\documentclass[preprint,prd,amsmath,amssymb,amsthm,nofootinbib]{revtex4}
\usepackage{xcolor}
\usepackage[mathscr]{euscript}
\usepackage{bm}% bold math
\usepackage{graphicx}% Include figure files
\usepackage[caption=false]{subfig}
\usepackage[colorlinks=true,citecolor=blue,linkcolor=red,urlcolor=blue]{hyperref}
\usepackage{ulem}

\newcommand{\bea}{\begin{eqnarray}}
\newcommand{\eea}{\end{eqnarray}}
\newcommand{\beq}{\begin{equation}}
\newcommand{\eeq}{\end{equation}}

\begin{document}

\title{Entanglement harvesting in cosmic string spacetime}
\author{Ying Ji$^{1}$, Jialin Zhang$^{1,2}$~\footnote{Corresponding author. jialinzhang@hunnu.edu.cn} and Hongwei Yu$^{1,2}$~\footnote{Corresponding author. hwyu@hunnu.edu.cn}}
\affiliation{
$^1$ Department of Physics and Synergetic Innovation Center for Quantum Effects and Applications, Hunan Normal University, 36 Lushan Rd., Changsha, Hunan 410081, China\\
$^2$ Institute of Interdisciplinary Studies, Hunan Normal University, 36 Lushan Rd., Changsha, Hunan 410081, China}
\date{\today}
\begin{abstract}
We investigate the entanglement harvesting phenomenon for static detectors that locally interact with massless scalar fields in the cosmic string spacetime, which, though locally flat, features a conical structure defined by a deficit angle. Specifically, we analyze three detector alignments relative to the string: parallel and orthogonal alignments with detectors on the same side of the string, and an orthogonal alignment with detectors on opposite sides of the string.
For the alignments on the same side of the string, we observe that the cosmic string's presence can either aid or hinder entanglement harvesting, affecting both the extent of entanglement harvested and the achievable range of interdetector separation. This effect depends on the distance between the detectors and the string and differs markedly from scenarios in a locally flat spacetime with a reflecting boundary, where the boundary invariably extends the harvesting-achievable range. Conversely, for the alignment with detectors on opposite sides of the string, we find that detectors consistently harvest more entanglement than those in a flat spacetime devoid of a cosmic string. This starkly contrasts the behavior observed with detectors on the same side. Interestingly, the presence of a cosmic string expands the harvesting-achievable range for detectors in orthogonal alignment only when near the string, whereas it invariably reduces the achievable range for detectors in parallel alignment.
\end{abstract}

\maketitle

\section{Introduction}

 In the framework of formal algebraic quantum field theory,  it has long been recognized  that the vacuum state of a free quantum field can maximally violate Bell's inequalities~\cite{Summers:1987,Reznik:2005},  indicating the presence of quantum entanglement between both timelike and spacelike separated regions within the vacuum state. This entanglement suggests that the vacuum itself may serve as a potential resource for quantum technologies.   Building on the foundational work of Valentini \cite{Valentini:1991} and later developments by Reznik \cite{Reznik:2003},  the concept emerged that vacuum entanglement could be extracted  through local interactions of multiple detectors with the field.
 This led to the formulation of what is now known as the entanglement harvesting protocol, wherein two initially uncorrelated Unruh-DeWitt (UDW) detectors interact locally with a quantum field (typically in the vacuum state) to extract entanglement \cite{Salton:2015, Kerstjens:2015}.   The phenomenon of entanglement harvesting has since been explored across various settings~\cite{Steeg:2008,Menicucci:2012,Hu:2012,Nambu:2013,Menicucci:2014,Smith:2015,Ng:2018-1,Henderson:2018,Ng:2018-2,
 Henderson:2019,CW:2019,Koaga:2019,CW:2020,Zhang:2020,Henderson:2020,Xu:2020,Liu:2021,Majhi:2021-1,Tjoa:2021,Maeso:2022,Zhjl:2022.4,Signgh:2023,Liu:2023}.
 Entanglement harvesting has been shown to be highly sensitive  to several aspects of spacetime,  including its topology~\cite{Smith:2015,Liu:2021,Liu:2023} and curvature~\cite{Menicucci:2014,Ng:2018-1,Henderson:2018,Ng:2018-2,Henderson:2019}, and those of the detectors, such  as their superpositions of temporal order~\cite{Henderson:2020},  intrinsic motion~\cite{Koaga:2019,Zhang:2020,Liu:2021,Majhi:2021-1} and  energy gaps~\cite{Maeso:2022,Zhjl:2022.4,Liu:2023}.   It has been argued that this sensitivity to topology can serve as a tool to differentiate between locally flat spacetimes that are distinct only in their topological structures \cite{Smith:2015}.
 Recent studies have uncovered that the presence of a reflecting boundary in a flat spacetime, effectively rendering the spacetime topologically nontrivial, plays a notable role in the dynamics of entanglement harvesting. Specifically, it has been found to inhibit entanglement harvesting close to the boundary while facilitating it further away \cite{Liu:2021, Liu:2023}.

 In addition to locally flat spacetime with a reflecting boundary, another fascinating example of a locally flat yet topologically nontrivial spacetime is characterized by a conical structure with a deficit angle. This structure is physically interesting as it describes the spacetime around a cosmic string, a type of topological defect \cite{Vilenkin:1981}. Cosmic strings may arise from phase transitions in the early universe \cite{Kibble:1976, Linde:1979} or within certain gauge extensions of the standard model of particle physics \cite{Vilenkin:1994}.
  The conical structure surrounding a cosmic string leads to a variety of intriguing cosmological, astrophysical, and gravitational phenomena \cite{Vilenkin:1994, Hindmarsh:1995, Copeland:2011}. This unique structure also influences quantum fluctuations of fields, which results in significant modifications to several quantum phenomena. Notable effects include alterations to the Casimir-Polder effect \cite{Saharian:2011, Saharian:2012}, atomic transitions \cite{Iliadakis:1995, Bilge:1998, Cai:2015, Zhou:2016}, resonance interactions \cite{Zhou:2018}, fluctuations of the lightcone \cite{Mota:2016}, and the dynamics of entanglement \cite{He:2020}.
  Noteworthily, there has been a proposal for the experimental detection of analogous spacetime metric fluctuations through the observable variance in flight time near fabricated analog cosmic strings, as detailed in Ref. \cite{Hu:2018}. This approach underscores the potential for experimental investigations into the effects of cosmic strings and similar topological features within controlled settings.

The simplest model of cosmic string spacetime features a deficit planar angle around a static, straight, and infinitely thin string \cite{Vilenkin:1981}. Previous studies have shown that the atomic transition rate and resonance interactions of atoms in this cosmic string spacetime can exhibit behaviors similar to those observed in a flat spacetime with a perfectly reflecting boundary \cite{Cai:2015, Zhou:2016, Zhou:2018}. Inspired by these findings, a natural question arises regarding the role of the cosmic string in the context of entanglement harvesting.  Since both the effects  of a reflecting plane boundary and a cosmic string on the vacuum fields can be studied  by considering the contributions from the ``images" due to the boundary and the conical structure of the string in the Wightman functions of the fields,
it is also quite interesting to compare the phenomenon of entanglement harvesting in the cosmic string spacetime with that in a flat spacetime with a reflecting boundary. Such  a comparison may provide a useful method to distinguish locally flat spacetimes that differ only in topology caused by cosmic strings and reflecting boundaries. These are what we are planning to explore in the present paper.

This paper is organized as follows. In the following section, we review  the entanglement harvesting protocol,  including the UDW detector model and the basic formula for  the detector-field coupling.  In Sec. \ref{sec3}, we  derive  the expressions for the detectors' transition probabilities and their nonlocal correlations in the cosmic string spacetime,  and investigate the entanglement harvesting phenomenon for two static detectors in three different alignments with respect to the cosmic string in detail. Numerical evaluations are employed when necessary  to clearly exhibit  the  behaviors of entanglement harvesting and comparisons are  made between the entanglement harvesting phenomena in the cosmic string spacetime and flat spacetime with a reflecting boundary.  Finally, we end  with summaries in Sec. \ref{sec4}.  For convenience, the natural units $\hbar=c=k_B=1$ are adopted throughout this paper.

\section{The basic formulas}
\label{sec2}
In the standard entanglement harvesting protocol,  one considers two UDW detectors $A$ and $B$ which locally interact with a massless quantum scalar field
$\phi[x_D(\tau)]$ ($D \in \{A, B\}$) along their worldlines. The classical  trajectory of the detector, $x_D(\tau)$, is  parameterized in terms
of its proper time $\tau$. Suppose that the UDW detector has an energy gap  $\Omega_D$  between its ground state $|0\rangle_{D}$ and excited state $|1\rangle_{D}$. Then the interaction Hamiltonian for such a detector locally coupling with the scalar field  in the interaction picture is given by
\begin{equation}\label{HD1}
H_{D}(\tau)=\lambda \chi(\tau)\big[e^{i \Omega_{D} \tau} \sigma^{+}+e^{-i \Omega_{D} \tau} \sigma^{-}\big] \phi\big[x_{D}(\tau)\big]\;,
\end{equation}
where constant $\lambda\ll1$ denotes the weak coupling strength, $\chi(\tau)=\exp\big[-\tau^2/(2\sigma^2_{D})\big]$ is the Gaussian switching function  that allows  to control the interaction duration  via the parameter $\sigma_D$, and $\sigma^{+}=|1\rangle_{D}\langle0|_{D}$ and $\sigma^{-}=|0\rangle_{D}\langle1|_{D}$ denote ladder operators acting on the Hilbert space of the detectors.

The initial state of the detector-field system, assuming that both Unruh-DeWitt (UDW) detectors
 $A$ and $B$  are prepared in their ground states and the scalar field is in its vacuum state $|0\rangle$, can be expressed as  $|\Psi_i\rangle=|0\rangle_{A}|0\rangle_{B}|0\rangle$.  During the interaction between the detectors and the field, the state evolves according to the dynamics dictated by the interaction Hamiltonian. This interaction can cause the detectors to become entangled, even though they do not interact directly with each other but only through the quantum field. The final state of the system, after the interaction has taken place, can be  shown to be given by
\begin{equation}
\left|\Psi_{f}\right\rangle:=\mathcal{T} \exp \Big[-i \int d t\Big(\frac{d \tau_{A}}{d t} H_{A}(\tau_{A})+\frac{d \tau_{B}}{d t} H_{B}(\tau_{B})\Big)\Big]|\Psi_i\rangle\;,
\end{equation}
where $\mathcal{T}$ denotes the time ordering operator and $t$ is the  coordinate time with respect to which
the vacuum state of the field is defined. For simplicity, we presume that the two detectors have identical  energy gaps $\Omega$ ($\Omega_{A}=\Omega_{B}$) and  interaction duration parameters parameter $\sigma$ ($\sigma_{A}=\sigma_{B}$).  Tracing out the field degrees of freedom,   one can obtain, using the perturbation theory, that the density matrix for the final state of the detectors in the basis ${|0\rangle_{A}|0\rangle_{B}, |0\rangle_{A}|1\rangle_{B}, |1\rangle_{A}|0\rangle_{B}, |1\rangle_{A}|1\rangle_{B}}$ is, to the leading order in the coupling
strength, given by~\cite{Smith:2015,Henderson:2018,Henderson:2019}
\begin{align}\label{rho}
\rho_{A B}: & =\operatorname{tr}_{\phi}\big(|\Psi_{f}\rangle\langle\Psi_{f}|\big)\nonumber\\
& =\left(\begin{array}{cccc}
1-P_{A}-P_{B} & 0 & 0 & X \\
0 & P_{B} & C & 0 \\
0 & C^{*} & P_{A} & 0 \\
X^{*} & 0 & 0 & 0
\end{array}\right)+\mathcal{O}(\lambda^{4})\;,
\end{align}
where
\begin{equation}\label{ppd}
P_{D}:=\lambda^{2} \iint d \tau d \tau' \chi(\tau) \chi(\tau') e^{-i \Omega(\tau-\tau')} W\big(x_{D}(t), x_{D}(t')\big), \quad D \in\{A, B\}\;,
\end{equation}
\begin{equation}\label{X}
X:=-\lambda^{2} \iint d \tau d \tau'\chi(\tau) \chi(\tau') e^{-i \Omega(\tau+\tau')}\Big[\theta(t'-t) W\big(x_{A}(t), x_{B}(t')\big)+\theta(t-t') W\big(x_{B}(t'), x_{A}(t)\big)\Big]\;,
\end{equation}
\begin{equation}
C:=\lambda^{2} \iint d \tau d \tau' \chi(\tau) \chi(\tau') e^{-i \Omega(\tau-\tau')} W\big(x_{A}(t), x_{B}(t')\big)\;,
\end{equation}
with $W(x,x'):=\langle0|\phi(x)\phi(x')|0\rangle$ being the Wightman function of the  scalar field in vacuum, and $\theta(x)$ denoting the Heaviside step function. In particular, if the two detectors are at rest, we have  $t=\tau$ and  $t'=\tau'$.  Here, the matrix element $P_{D}$ represents the detector's transition probability from the ground state to the excited state due to the interaction between the detector and the field, and the quantities $C$ and $X$ represent the nonlocal correlations between the two detectors.

To quantify the entanglement acquired by the two detectors, we  employ  concurrence as a measure of entanglement~\cite{WW}.  For the density matrix~\eqref{rho}, the concurrence can be evaluated straightforwardly\cite{Smith:2015,Henderson:2018,Henderson:2019},
\begin{equation}\label{Con}
\mathcal{C}(\rho_{A B})=2 \max \Big[0,|X|-\sqrt{P_{A} P_{B}}\Big]+\mathcal{O}(\lambda^{4})\;.
\end{equation}
This indicates that  the concurrence $\mathcal{C}(\rho_{A B})$ is determined by the competition between  the nonlocal correlation term $X$ and the geometric mean of the transition probabilities $P_A$ and $P_B$, which both crucially depend on the Wightman function of the scalar field.

\section{Harvesting Entanglement in the cosmic string spacetime}
\label{sec3}
We now begin to study the entanglement harvesting phenomena for two UDW detectors  near a long straight cosmic string which is situated along the $z$-axis in a flat spacetime.  The line element of the cosmic spacetime  can be written in cylindrical coordinates as~\cite{Vilenkin:1981}
\begin{equation}\label{metr1}
ds^{2}=dt^{2}-d\rho^{2}-\rho^{2}d\theta^{2}-d z^{2}\;,
\end{equation}
where $\theta\in[0,2\pi/\nu]$ and $\nu:=(1-4G\mu)^{-1}$ with $G$  and $\mu$ being the Newton's gravitational constant  and the cosmic string linear energy density, respectively.
The dimensionless quantity $G\mu$ measures the strength of the gravitational effects of the string manifesting as  a deficit angle $\delta\theta:=2\pi(\nu-1)/\nu=8\pi{G}\mu$ with respect to  the trivial flat spacetime (Minkowski spacetime). Notice that for  the cosmic string spacetime one always has the deficit-angle parameter $\nu>1$ .

\begin{figure}[!htbp]\centering
\includegraphics[width=0.65\textwidth]{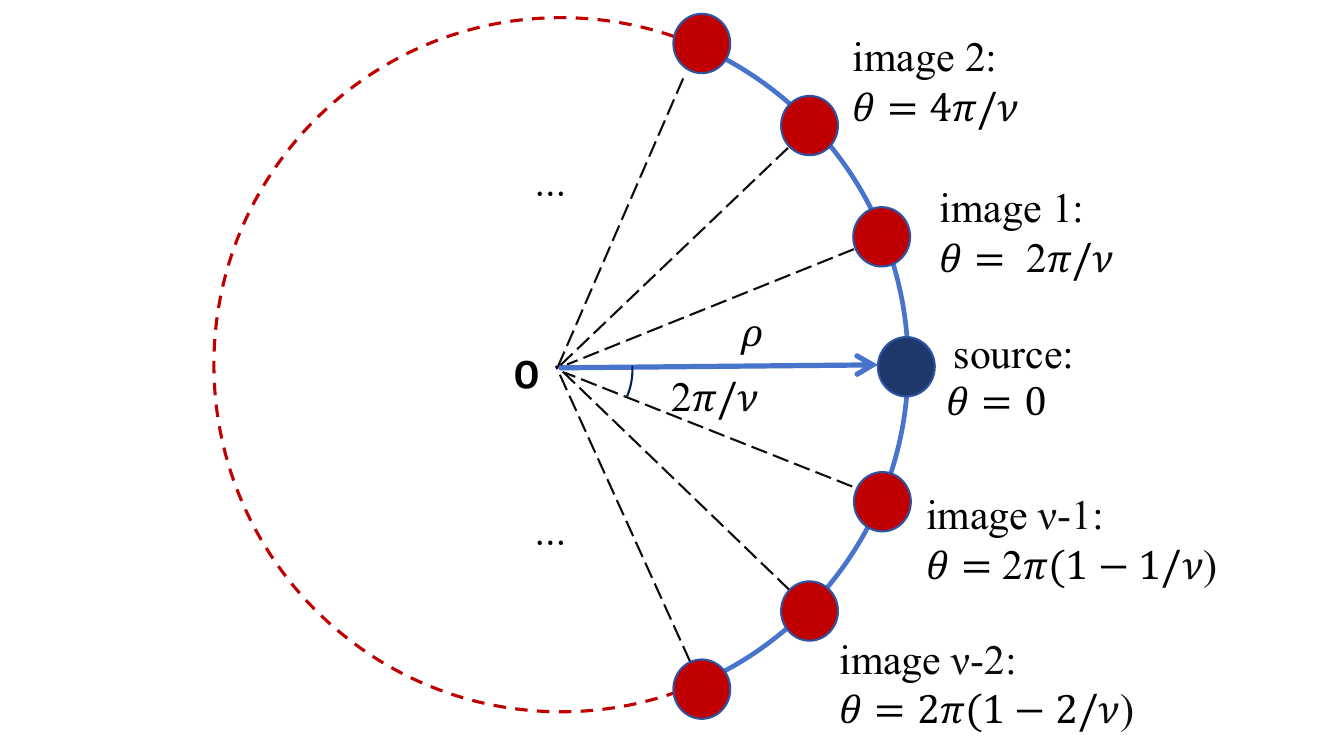}
\caption{The  ``images" for integer $\nu$ due to the conical topology with a deficit angle $2\pi(\nu-1)/\nu$.}\label{vimages}\end{figure}

After analytically solving  the  Klein-Gordon equation for a massless scalar field,
the Wightman function of the field in the cosmic string spacetime can  be obtained~\cite{Mota:2016}
\begin{align}\label{wxx}
W\big(x, x'\big) =&\frac{1}{4\pi^{2}}\frac{1}{\sigma_{0}^{2}}+\frac{1}{2\pi^{2}}\sum_{m=1}^{[\nu/2]}{'}\frac{1}{\sigma_{m}^{2}}\nonumber\\
&-\frac{\nu}{8\pi^{3}}\sum_{j=+,-} \int_{0}^{\infty} d\zeta\frac{\sin(j\nu\Delta\theta+\nu\pi)}{[\cosh(\nu\zeta)-\cos (j\nu\Delta\theta+\nu\pi)]} \frac{1}{\sigma_{\zeta}^{2}}\;,
\end{align}
where
\begin{align}
\sigma_{0}^{2} & = -\Delta t^{2}+\Delta z^{2}+\rho^{2}+\rho^{\prime 2}-2\rho\rho^{\prime} \cos \Delta\theta , \\
\sigma_{m}^{2} & = -\Delta t^{2}+\Delta z^{2}+\rho^{2}+\rho^{\prime 2}-2\rho\rho^{\prime} \cos \Big(\frac{2\pi m}{\nu}-\Delta \theta\Big) , \\
\sigma_{\zeta}^{2} & = -\Delta t^{2}+\Delta z^{2}+\rho^{2}+\rho^{\prime 2}+2\rho\rho^{\prime}\cosh\zeta ,
\end{align}
with $\Delta t=t-t'-i\epsilon$, $\Delta z=z-z'$, and $\Delta \theta=\theta-\theta'$. Here, $[\nu/2]$ denotes the integer part of  $\nu/2$ and the prime in the summation means that when $\nu$ is an even integer the term with $m=\nu/2$  should be multiplied by an additional factor 1/2. Obviously,   the summation has no contribution when  $\nu<2$, and the Wightman function~(\ref{wxx}) is generally discontinuous as a function  of $\nu$ due to  the integer truncation  operation $[\nu/2]$.
Furthermore, if  $\nu$ is an integer number, the third term in Eq.~(\ref{wxx}) will vanish and the Wightman function takes a simple form
\begin{align}\label{wxx2}
W\big(x, x'\big) =&\frac{1}{4 \pi^{2}} \frac{1}{\sigma_{0}^{2}}+\frac{1}{4 \pi^{2}} \sum_{m=1}^{\nu-1}{} \frac{1}{\sigma_{m}^{2}}\;.
\end{align}
The first term in the above equation is just the Wightman function in a trivial flat spacetime  and the second term is a summation of  the Wightman functions
corresponding to  $\nu-1$   ``images"  due to the conical topology with a planar deficit angle [see Fig.~(\ref{vimages})]. Obviously, as the distance from the cosmic string increases, the spatial separation between the source and its images also grows larger. Consequently, the contributions of these images to the Wightman function become negligible in the regions far from the cosmic string.

\begin{figure}[!htbp]
\centering
\subfloat[]{\label{model1}\includegraphics[width=0.325\linewidth]{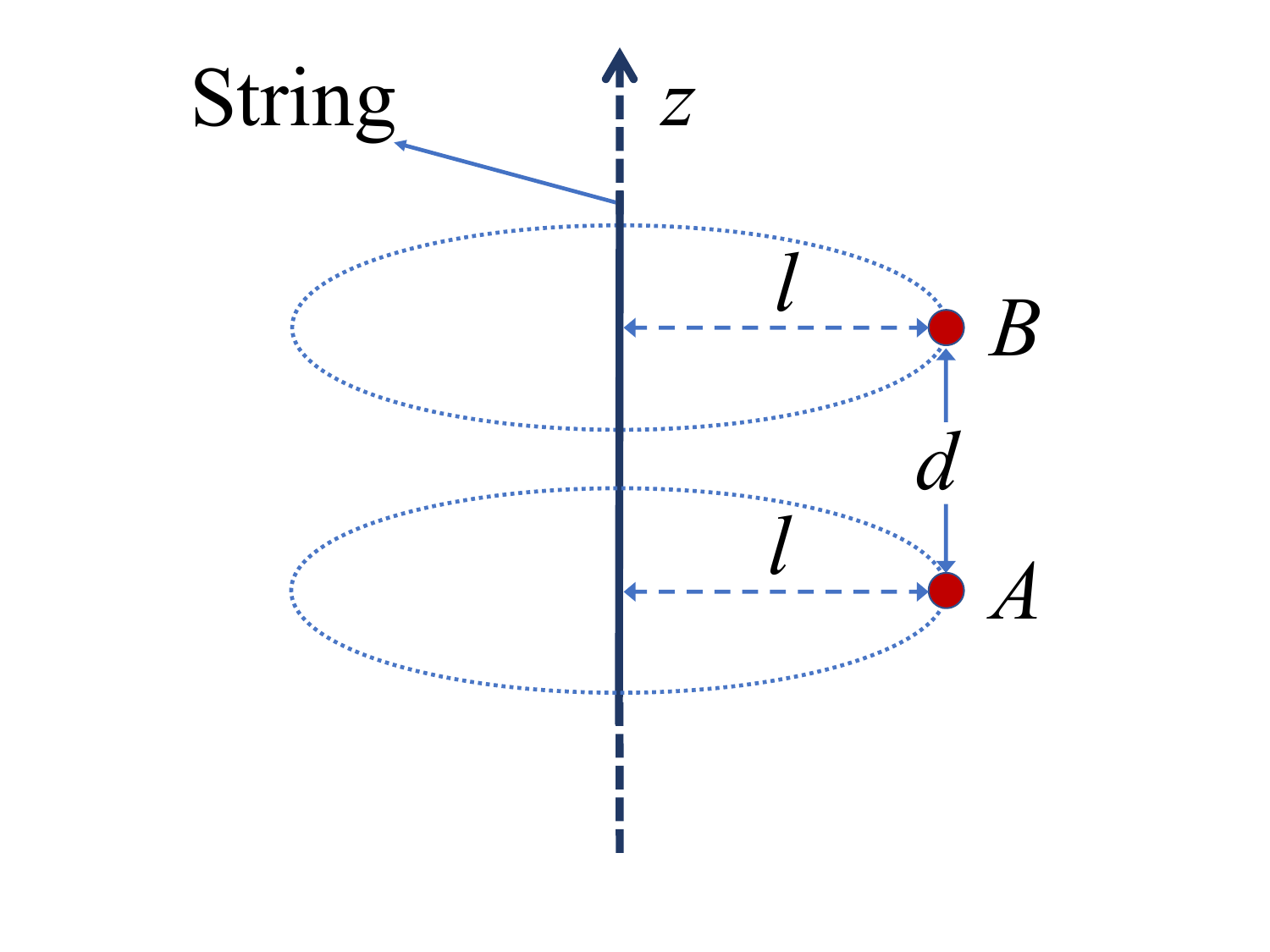}}
\subfloat[]{\label{model2}\includegraphics[width=0.325\linewidth]{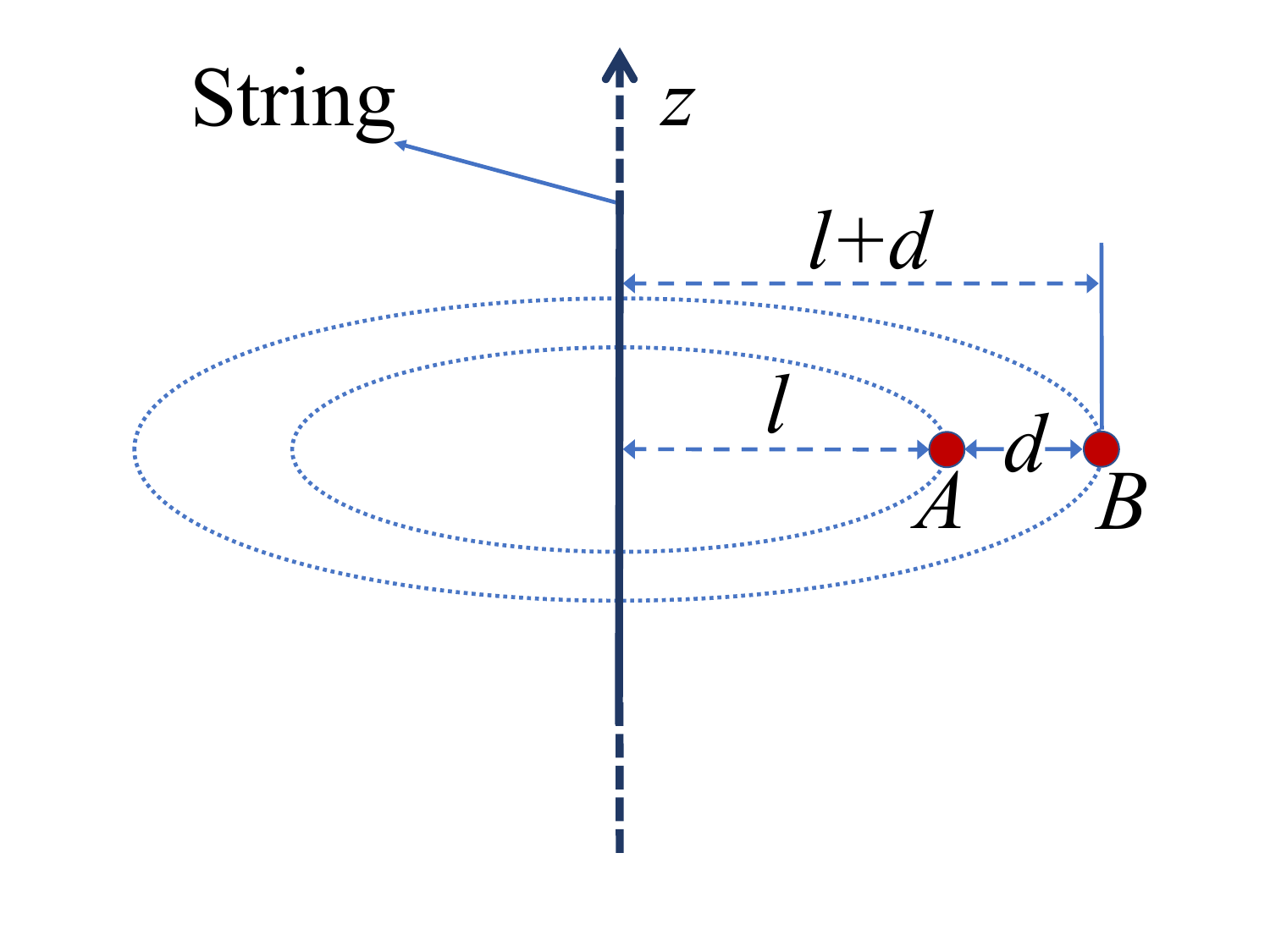}}
\subfloat[]{\label{model3}\includegraphics[width=0.325\linewidth]{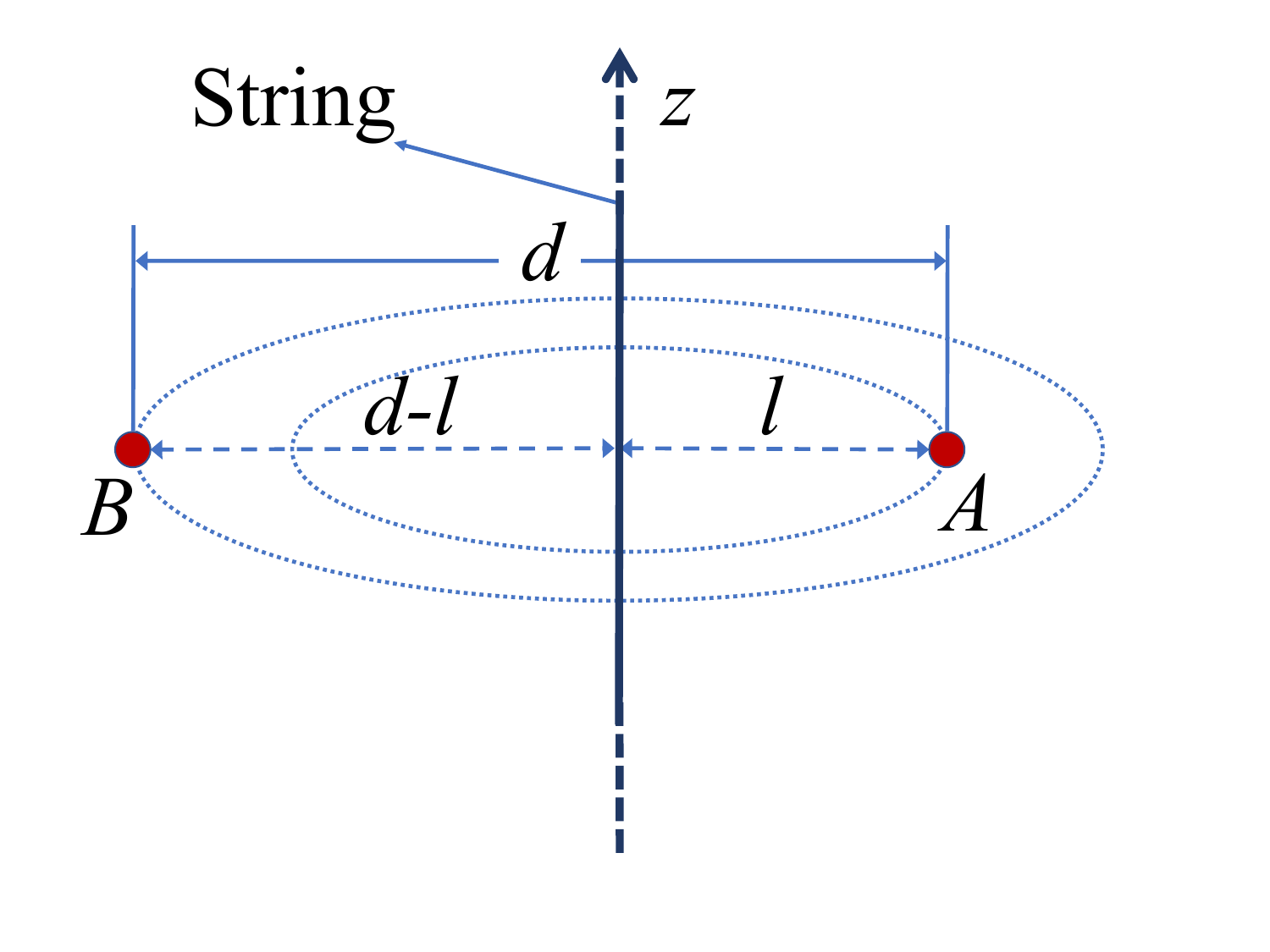}}
\caption{ The cosmic string is assumed to be located along the $z$-axis. Plots (a) and (b) respectively describe the interdetector separation $d$  aligned  parallel and orthogonally  to the cosmic string on the same side of the string.  Plot (c) shows that the two detectors are  orthogonally aligned  to the string on  two opposite sides. }\label{model}
\end{figure}

Without loss of generality, we now  assume  that  two static UDW detectors,  separated by a distance $d$, are positioned relative to the string in three different alignments: the line connecting the detectors is parallel to the string,  orthogonal  to the string with two detectors on the same side and two detectors respectively on two  opposite sides, as illustrated in Figure~(\ref{model}).

\subsection{Alignments on the same side of the string}

\subsubsection{Parallel alignment}
We start with the parallel alignment [see Fig. \eqref{model1}].   The spacetime trajectories for the two detectors  in this case can be expressed as
\begin{equation}\label{para-x}
x_{A}:= \left\{t=\tau,~\rho=l,~\theta=0,~z=0\right\},~x_{B}:= \left\{t=\tau,~\rho=l,~\theta=0,~z=d\right\}\;.
\end{equation}
In order to calculate the concurrence, we  first need to calculate the transition probabilities of the detectors.
Substituting the trajectory~(\ref{para-x})  into Eq.~(\ref{wxx}) yields
 \begin{align}\label{WW2}
W\big(x_D, x_D'\big)= & -\frac{1}{4 \pi^{2}} \frac{1}{(\tau-\tau'-i\epsilon)^2}-\frac{1}{2 \pi^{2}} \sum_{m=1}^{[\nu/2]}{'} \frac{1}{(\tau-\tau'-i\epsilon)^2-4l^2\sin^2({m\pi}/{\nu})} \nonumber\\
& +\frac{\nu}{4 \pi^{3}} \int_{0}^{\infty} d \zeta \frac{\sin (\nu\pi)}{[\cosh (\nu \zeta)-\cos (\nu \pi)][(\tau-\tau'-i\epsilon)^2-4l^2\cosh^2({\zeta}/{2})]}\;.
\end{align}
After some mathematical manipulations [see Appendix \ref{appd1}], the transition probability~(\ref{ppd}) can be expressed as follows
\begin{equation}\label{pd}
P_{D}=P_{0}+P_{1}+P_{2},\quad D \in\{A, B\}\;
\end{equation}
with
\begin{equation}
P_{0}=\frac{\lambda^{2}}{4\pi}\left[e^{-\sigma^{2}\Omega^{2}}-\sqrt{\pi}\sigma\Omega \operatorname{Erfc}(\sigma \Omega)\right]\;,
\end{equation}
\begin{align}
P_{1}=&\frac{\lambda^{2} \sigma}{4 \sqrt{\pi}}\sum_{m=1}^{[\nu/2]}{'}\;\frac{e^{-l^{2}\sin^{2}({m\pi}/{\nu})/\sigma^{2}}}{l\sin({m\pi}/{\nu})}\bigg\{\operatorname{Im}\Big[e^{2 i \Omega l\sin({m\pi}/{\nu})} \operatorname{Erf}\Big(\frac{i l}{\sigma}\sin\frac{m\pi}{\nu}+\sigma\Omega\Big)\Big]\nonumber\\
&-\sin \Big[2 \Omega l\sin\big({m\pi }/{\nu}\big)\Big]\bigg\}\;,
\end{align}
\begin{align}
P_{2}=&\frac{\lambda^{2} \sigma\nu\sin(\nu\pi)}{8 \pi^{3/2}}\int_{0}^{\infty} d\zeta\frac{1}{\cosh(\nu\zeta)-\cos(\nu\pi)} \frac{e^{-l^{2} \cosh^{2}({\zeta}/{2})/\sigma^{2}}}{l \cosh({\zeta}/{2})}\bigg\{\sin\big[2l\Omega  \cosh({\zeta}/{2})\big]\nonumber\\
&-\operatorname{Im}\Big[e^{2 i l\Omega  \cosh({\zeta}/{2})} \operatorname{Erf}\Big(\frac{{i}l}{\sigma}\cosh\frac{\zeta}{2}+\sigma \Omega\Big)\Big]\bigg\}\;,
\end{align}
where $\operatorname{Erfc}(x):=1-\operatorname{Erf}(x)$ with the error function being $\operatorname{Erf}(x):=\int_{0}^{x} 2e^{-t^{2}}d t/\sqrt{\pi}$.

Note that the first term $P_{0}$ is just  the transition probability  for a static detector in a flat spacetime without a cosmic string, which  approaches zero in the limit of $\Omega\sigma\rightarrow\infty$ as expected~\cite{Smith:2015}. Obviously, the transition probability $P_D$ reduces to $P_{0}$  in the limit of $l\rightarrow\infty$  or $\nu=1$, i.e.,  when the detectors are located infinitely far from the string, or when the deficit angle vanishes. Moreover, for small $l/\sigma$, namely, when the detectors are very close to the string, the transition probability $P_D$ can  be further  approximated as
\begin{equation}\label{plsmall}
P_{D}\approx{\nu{P_{0}}}=\frac{\lambda^{2}\nu}{4\pi}\left[e^{-\sigma^{2}\Omega^{2}}-\sqrt{\pi}\sigma\Omega \operatorname{Erfc}(\sigma \Omega)\right]\;.
\end{equation}
So, the transition probability of the detector in the presence of a cosmic string is $\nu$ times that  in  a flat spacetime without a cosmic string, suggesting that the presence of the deficit angle  increases the detector's transition probability. Especially, when the detector is located on the string, the transition probability can be expressed in a particularly simple form
\begin{align}\label{L0}
P_{D}&=P_0+2\sum_{m=1}^{[\nu/2]}{'}\;P_{0}+\frac{\nu}{\pi}\int_{0}^{\infty} d \zeta \frac{\sin (\nu \pi)}{\cos (\nu \pi)-\cosh (\nu \zeta)}{P_{0}}\nonumber\\
 &=\left\{\begin{aligned}
&\nu{P_0},&{\text{integer}}~\nu\\
&\Big(1+2\cdot\big[\frac{\nu}{2}\big]-\frac{2}{\pi}\operatorname{arctan~cot}\frac{\nu\pi}{2}\Big)P_0,&{\text{non-integer}}~\nu\;
 \end{aligned} \right.
\nonumber\\ &=\nu{P_0}\;,
\end{align}
where in the last step we have used the fact that for non-integer $\nu$, the expression simplifies as follows
$$-\frac{2}{\pi}\operatorname{arctan~cot}\frac{\nu\pi}{2}=\frac{i\ln(-e^{-i\pi\nu})}{\pi}=\frac{i}{\pi}\ln{e}^{-i\pi(\nu-1)}=\frac{i}{\pi}\ln{e}^{-i\pi(\nu-1-2\cdot[\nu/2])}=\nu-1-2\cdot\big[\frac{\nu}{2}\big]$$  with $\big|\nu-1-2\cdot[\nu/2]\big|<1$ ensured by the requirement of  the principal value of the logarithmic function.
From the $l$-dependence  in $P_1$ and $P_2$, one can also infer that the transition probability  is a monotonically decreasing function of the detector-to-string distance $l$, with the maximum value attained at $l=0$ for a fixed $\nu$.

\begin{figure}[!htbp]
\centering
\subfloat[]{\label{P-L}\includegraphics[width=0.48\linewidth]{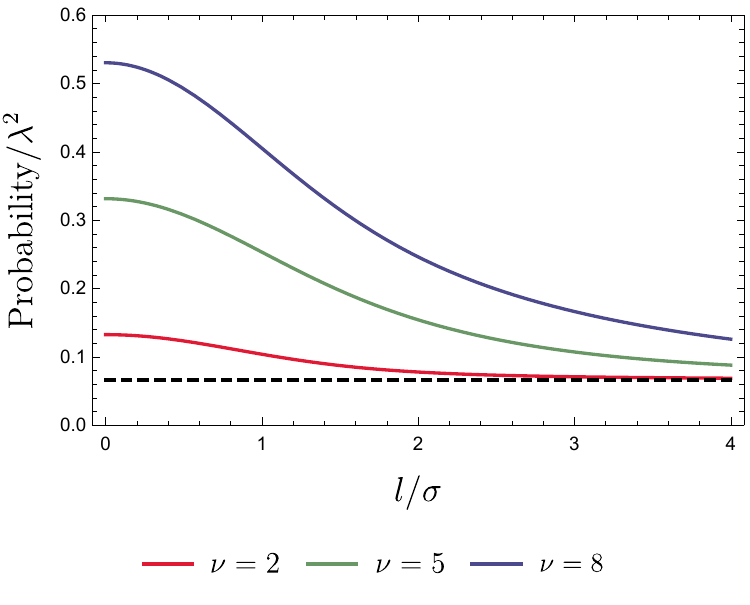}}\quad
\subfloat[]{\label{P-nu}\includegraphics[width=0.48\linewidth]{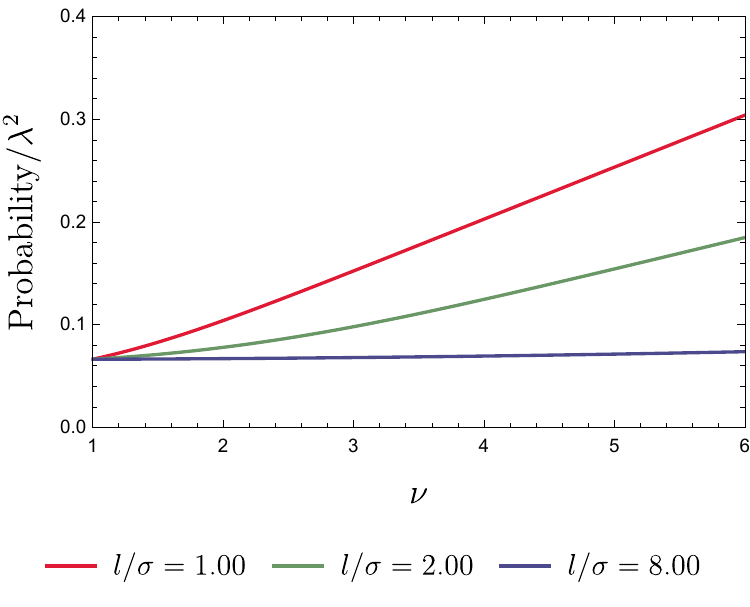}}
\caption{The transition probability is plotted as a function of  $l/\sigma$  in (a) and as a function of $\nu$ in  (b) with fixed parameter $\Omega\sigma=0.10$.
 %In all plots,we have set $\nu=\{2,5,8\}$.
Here, the black dashed line in the left plot represents the corresponding results in a flat spacetime without a cosmic string (i.e., $\nu=1$). For convenience, all relevant physical quantities are expressed in the unit of the interaction duration parameter $\sigma$.}\label{P-L-nu}
\end{figure}

In order to clearly show  how  the transition probability  depends on the detector-to-string distance and the deficit angle, we have resorted to numerical calculations. The results are presented in Fig.~(\ref{P-L-nu}).
Obviously, the transition probability in the cosmic string spacetime generally decreases as  the detector-to-string distance grows, ultimately converging  to the  result observed in a flat spacetime  without a cosmic string when the detector-to-string distance approaches  infinity [see Fig.~(\ref{P-L})]. Moreover, the transition probability indeed reaches its maximum when the detector is positioned directly on the string, confirming our earlier analytical analysis. It is also easy to  see that the transition probability is an  increasing function of  the  deficit-angle parameter $\nu$. This means that  the  deficit angle  enhances the transition probability [see Fig.~(\ref{P-nu})].

Let us now turn to calculate the correlation term $X$. For convenience, we denote it by $X_{P}$ for the case of the parallel alignment. Substituting  the trajectories~(\ref{para-x}) and the Wightman function~(\ref{wxx}) into  Eq.~(\ref{X}) , we have [see Appendix \ref{appd2}]
\begin{equation}\label{Xp}
X_{P}=X_0+X_{P1}+X_{P2}\;
\end{equation}
with
\begin{equation}
X_0=f\Big(\frac{d}{2\sigma}\Big)\;,
\end{equation}
\begin{equation}
X_{P1}=2\sum_{m=1}^{[\nu/2]}{'}f\Big(\sqrt{\frac{d^2}{4\sigma^2}+\frac{l^2}{\sigma^2}\sin^{2}\frac{m\pi}{\nu}} \Big)\;,
\end{equation}
and
\begin{equation}
X_{P2}=\int_{0}^{\infty}\mathrm{d}\zeta\frac{\nu\sin(\nu\pi)}{\pi[\cos(\nu\pi)-\cosh(\nu\zeta)]}
f\Big(\sqrt{\frac{d^2}{4\sigma^2}+\frac{l^2}{2\sigma^2}+\frac{l^2\cosh\zeta}{2\sigma^2}} \Big)\;,
\end{equation}
where the auxiliary function $f$ is defined as
\begin{equation}\label{fzd}
f(z):=-\frac{i \lambda^{2}e^{-\sigma^{2}\Omega^{2}-z^2} }{8 \sqrt{\pi}} \frac{\operatorname{Erfc}(iz)}{z}\;.
\end{equation}
From the above equations, one can see that the correlation term $X_P$ will  be exponentially suppressed when the detector energy gap $\Omega$ is  increased.  Thus, a large energy gap  still plays a strong inhibitory role in entanglement harvesting.

Notice that the first term of the correlation  term $X_P$,
\begin{equation}
X_{0}=-\frac{i \lambda^{2} \sigma}{4 \sqrt{\pi}} \frac{\operatorname{Erfc}\left[id/(2\sigma)\right]}{d}e^{-\sigma^{2}\Omega^{2}-\frac{d^{2}}{4 \sigma^{2}}},
\end{equation}
is just the result for  a flat spacetime without a cosmic string. It is worth noting that  the  magnitude of $|X_{0}|$   diverges in the limit of $d\rightarrow0$. This divergence is due to the ill-defined point-like approximation of the UDW detector model for $d/\sigma\ll\lambda$, resulting in a mathematically divergent concurrence.  However,  it has been shown that a finite-size detector model with a spatial smearing function  can resolve this divergence issue in the entanglement harvesting protocol~\cite{Kerstjens:2015}. The second term $X_{P1}$ and the third term $X_{P2}$  are dependent on the detector-to-string distance $l$, and both of them  vanish  in the limit of $l\rightarrow\infty$, i.e., when the detectors  are very far away from the string.   As a result,    $X_{P}$ reduces to that in a flat spacetime without a cosmic string, $X_{0}$, and so does the concurrence.
While, when the detectors are very close to the string, i.e., when  $l/\sigma\ll1$, the correlation term can be approximated as
\begin{equation}\label{xlsmall}
X_{P}\approx\nu X_{0}-\nu\Big(\frac{l^2}{d^2}+\frac{l^2}{2\sigma^2}\Big)X_{0}-\frac{e^{-\sigma^{2}\Omega^{2}}l^{2}\lambda^{2}\nu}{4\pi d^{2}}\;.
\end{equation}
Then using Eqs.~(\ref{Con}),~(\ref{plsmall}) and~(\ref{xlsmall}), one finds, for the concurrence quantifying the entanglement harvested by the detectors,
\begin{equation}\label{Clsmall}
\mathcal{C}_{P}(\rho_{A B})\approx\nu \mathcal{C}_{0}(\rho_{AB})\;
\end{equation}
with \begin{equation}
\mathcal{C}_{0}(\rho_{AB})=\max\Big\{\frac{\lambda^{2}e^{-\sigma^{2}\Omega^{2}}}{2\sqrt{\pi}}\Big[\frac{\sigma}{d}e^{-d^{2}/(4\sigma^{2})}\Big|\operatorname{Erfc}\Big( \frac{id}{2\sigma}\Big) \Big|+e^{\sigma^{2}\Omega^{2}}\sigma \Omega \operatorname{Erfc}(\sigma \Omega)-\frac{1}{\sqrt{\pi}} \Big],0\Big\}\end{equation}
representing the concurrence  in  the case of a flat spacetime without a cosmic string. In particular,  when the detectors are positioned on the string ($l=0$),  we have
\begin{align}\label{XPCPl0}
X_{P}=\nu{X_0},~\mathcal{C}_{P}(\rho_{A B})=\nu{\mathcal{C}_{0}(\rho_{A B})}\;.
\end{align}
This means that the presence of the cosmic string amplifies the amount of  entanglement harvested by the detectors in its vicinity. Furthermore, the bigger the deficit-angle parameter $\nu$, the greater  the concurrence $\mathcal{C}_{P}(\rho_{AB})$.  However, the analytical approximations of  $X_P$ and $\mathcal{C}_{P}(\rho_{AB})$ are only obtainable in the asymptotic regions, i.e., when detectors are very close or very far away from the string. For more general locations, numerical evaluations will be needed and performed later, following an analytical analysis of the orthogonal alignment case.

\subsubsection{Orthogonal alignment}
In the case of the orthogonal  alignment with both detectors  on the same side of the string [see Fig. \eqref{model2}], the spacetime trajectories  can be written as
\begin{equation}\label{vert-x}
x_{A}:= \left\{t=\tau,~\rho=l,~\theta=0,~z=0\right\},~x_{B}:= \left\{t=\tau,~\rho=l+d,~\theta=0,~z=0\right\}
\end{equation}
with $l$ representing the distance to the string of the detector which is closer.  It is easy to see that the transition probability for detector $A$  is just given by Eq.~(\ref{pd}), and $P_B$ can be obtained  from  the same equation by replacing $l$ with $l+d$. Similarly,  the correlation term $X$, denoted now by $X_{V}$ for the orthogonal  alignment, can be written in the form:
\begin{equation}\label{Xvv}
X_{V}=X_{0}+X_{V1}+X_{V2}\;
\end{equation}
with
\begin{equation}\label{Xvv1}
X_{V1}=2\sum_{m=1}^{[\nu/2]}{'}f\Big(\sqrt{\frac{d^2}{4\sigma^2}+\frac{l(l+d)}{\sigma^2}\sin^{2}\frac{m\pi}{\nu}} \Big)
\end{equation}
and
\begin{equation}\label{Xvv2}
X_{V2}=\int_{0}^{\infty}\mathrm{d}\zeta\frac{\nu\sin(\nu\pi)}{\pi[\cos(\nu\pi)-\cosh(\nu\zeta)]}
f\Big(\sqrt{\frac{d^2}{4\sigma^2}+\frac{l(l+d)}{2\sigma^2}+\frac{l(l+d)\cosh\zeta}{2\sigma^2}} \Big)\;.
\end{equation}
Analogous to the correlation term in the parallel alignment, $X_V$ also approaches $X_0$ in the limit of $l\rightarrow\infty$, and as a result, the concurrence becomes that in the case of a flat spacetime without  a cosmic string. Moreover, for small $l/\sigma$ and $d/\sigma$, the correlation term $X_{V}$  aligns with  Eq.~(\ref{xlsmall}), so that  the concurrence can be approximately written as $\mathcal{C}_{V}(\rho_{A B})\approx\nu \mathcal{C}_{0}(\rho_{A B})$.

After presenting the analytical analysis for the asymptotic regions, we now begin our numerical analyses for  general  locations of the detectors.
Here, it is worth pointing out  that when the energy gap of the detectors is much larger than the Heisenberg energy ($\Omega\sigma\gg1$), both the transition probability and the correlation term are vanishingly small so that entanglement can hardly be harvested. So, we will only consider a not-too-large energy gap in the following numerical evaluations.

\begin{figure}[!htbp]
\centering
\subfloat[$\nu=2$]{\label{Cd1}\includegraphics[width=0.48\linewidth]{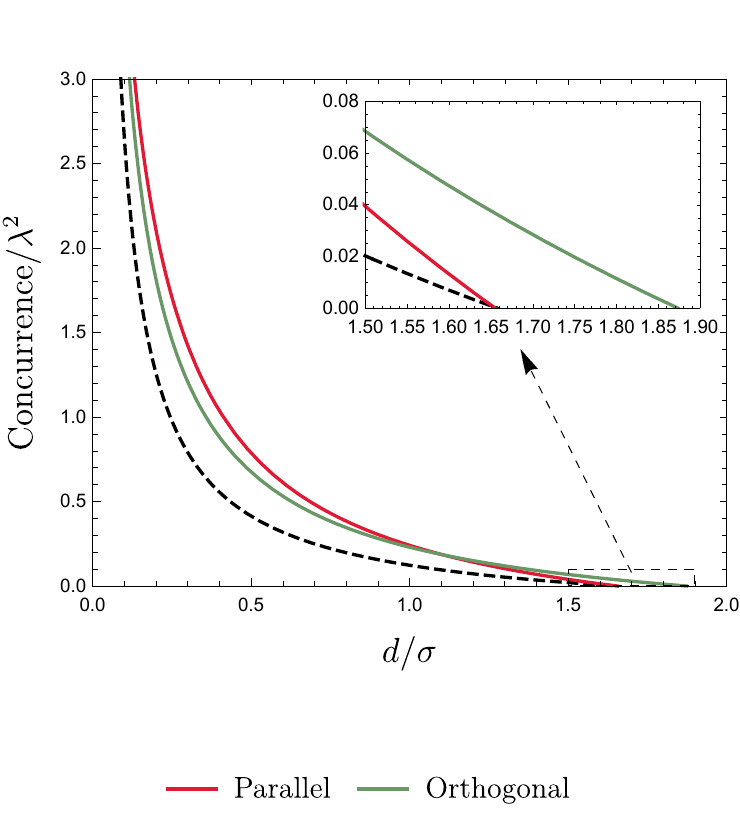}}\;
\subfloat[$\nu=11$]{\label{Cd3}\includegraphics[width=0.48\linewidth]{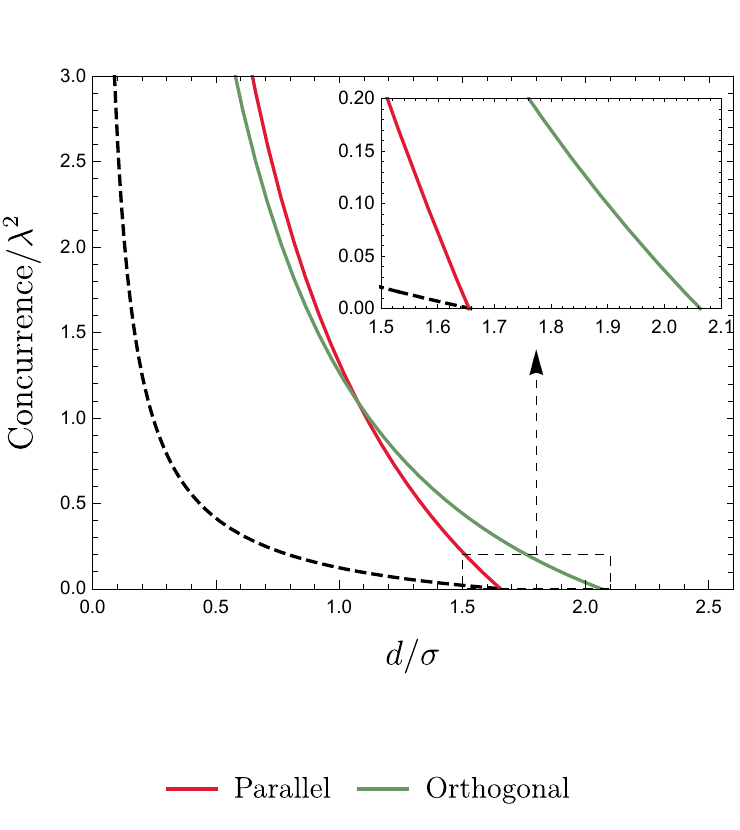}}
\caption{The concurrence versus $d/\sigma$ for $\nu=2$ in plot (a) and $\nu=11$ in plot (b) with $\Omega\sigma=0.10$ and $l/\sigma=0.10$. Here, the black dashed curves in all plots correspond to the results in a flat spacetime without a cosmic string.}\label{Cd}
\end{figure}
In Fig. \eqref{Cd}, we demonstrate how the concurrence varies with  interdetector separation for fixed detector-to-string distance (i.e., fixed $l/\sigma$) and deficit angle (characterized by parameter $\nu$).
Clearly, the concurrence decreases monotonically as the interdetector separation increases, regardless of the detectors' alignment relative to the cosmic string.  Interestingly, at small interdetector separations ($d/\sigma\ll1$), the two detectors aligned parallel to the string  will  harvest more entanglement than those in orthogonal  alignment for a fixed small $l/\sigma$, while at not too small interdetector separations, the detectors aligned orthogonally to the string will  instead harvest more entanglement.  One may  understand this property as follows.
When both the interdetector separation and detector-to-string distance are  small ($d/\sigma\ll1$ and $l/\sigma\ll1$), the transition probabilities of the detectors in these  two alignments, given by Eq.~(\ref{pd}), are almost the same. However,  comparing Eq.~(\ref{Xp}) with Eq.~(\ref{Xvv}), one can easily infer that the correlation term $|X_{P}|$ is larger than $|X_{V}|$ because the auxiliary function  $f(\cdot)$ is an  exponentially decreasing function of its argument. Therefore, one can see from Eq.~(\ref{Con}) that $\mathcal{C}_{P}(\rho_{AB})$ is larger than $\mathcal{C}_{V}(\rho_{AB})$ for small $d/\sigma$. When interdetector separation is  not too small ($d/\sigma>1$ and $d\gg{l}$), the correlation term  $|X_{P}|$ becomes approximately equal to $|X_{V}|$. However,  the geometric mean of the detectors' transition probabilities in the case of the orthogonal  alignment is much smaller than that in the parallel alignment. This is because the transition probability is a decreasing function of detector-to-string distance and the orthogonal alignment has a comparatively longer effective distance from the string than the parallel  alignment. As a result, $\mathcal{C}_{V}(\rho_{AB})$ is  greater than $\mathcal{C}_{P}(\rho_{AB})$ for not too small interdetector separations ($d/\sigma>1$ and $d\gg{l}$).

\begin{figure}[!htbp]
\centering
\subfloat[$\nu=2$]{\label{CL11}\includegraphics[width=0.48\linewidth]{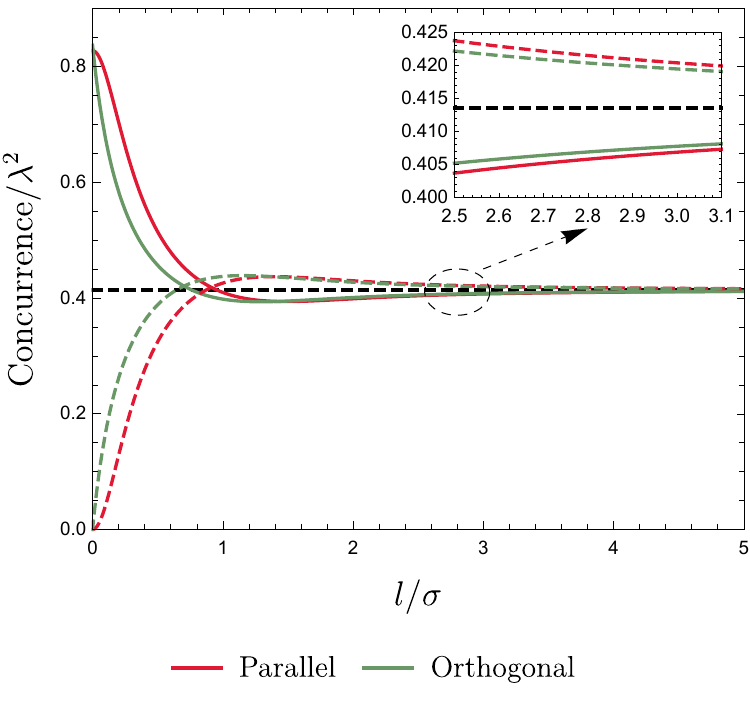}}\;
\subfloat[$\nu=11$]{\label{CL12}\includegraphics[width=0.48\linewidth]{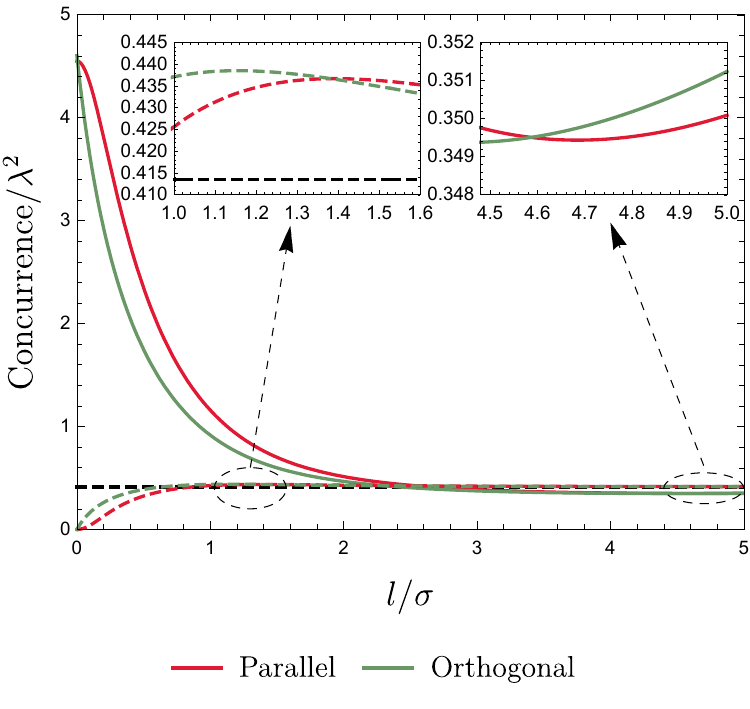}}
\caption{The concurrence is plotted as a function of $l/\sigma$ for $\nu=2$ in plot (a) and $\nu=11$ in plot (b) with fixed $\Omega\sigma=0.10$ and $d/\sigma=0.50$. The red dashed and green dashed lines respectively describe the corresponding cases of  parallel and orthogonal alignments in the presence of a reflecting boundary. The black horizontal dashed lines  indicate the corresponding results in a flat spacetime without any cosmic strings and boundaries.}\label{CL1}
\end{figure}

According to the aforementioned analysis, the harvested entanglement will approach  the result in a flat spacetime without a cosmic string as the detector-to-string distance grows to infinity.  This is analogous to the entanglement harvesting phenomenon for such detectors far way from the boundary in a flat spacetime with a reflecting boundary. To gain a better understanding of how  entanglement harvesting depends upon the detector-to-string distance, we plot  the concurrence as a function of detector-to-string distance for  various $\nu$ values in  Fig.~(\ref{CL1}). In addition,  the
dependence of concurrence on the detector-to-boundary distance in a flat spacetime with a reflecting plane boundary is also depicted in these plots for comparison.  To facilitate this comparison, we refer to the results for the case with a reflecting boundary  in a flat spacetime,  as detailed in references ~\cite{Liu:2021,Liu:2023}. The transition probability of the detector, according to these studies, satisfies [see Eq.~(3.4) in Ref.~\cite{Liu:2021}]
\begin{equation}\label{pdbd}
P_{D}^{bd}=P_0-\frac{\lambda^{2} \sigma e^{-l^{2}/\sigma^{2}}}{8 \sqrt{\pi}l}\bigg\{\operatorname{Im}\Big[e^{2 i l\Omega} \operatorname{Erf}\Big(\frac{i l}{\sigma}+\sigma\Omega\Big)\Big]-\sin \big(2l\Omega \big)\bigg\},\quad D \in\{A, B\}\;,
\end{equation}
where $l$ represents the distance between the detector and the boundary. Quite differently from the transition probability~(\ref{pd}) in the cosmic string spacetime, as described by Eq.~(\ref{pdbd}) does not reach a maximum value but instead vanishes when the detector is positioned directly at the boundary ($l=0$).  The correlation term $X$  in the case  of the parallel-to-boundary  alignment satisfies, for an interdetector separation $d$,
\begin{equation}\label{bdx1}
X_{P}^{bd}=X_{0}-f\Big(\sqrt{\frac{d^2}{4\sigma^2}+\frac{l^2}{\sigma^2}}\Big)\;,
\end{equation}
while in the case of  the orthogonal-to-boundary  alignment it becomes
\begin{equation}\label{xvbd}
X_{V}^{bd}=X_{0}-f\Big(\frac{d}{2\sigma}+\frac{l}{\sigma}\Big)\;
\end{equation}
with  $l$ now being  the distance to the boundary of the detector which is closer~[see Ref.~\cite{Liu:2023} for more details].
 Accordingly, the concurrence in a flat spacetime with a boundary  can be straightforwardly obtained by substituting  Eqs.~(\ref{pdbd})~(\ref{bdx1}) and (\ref{xvbd}) into Eq.~(\ref{Con}). In comparing Eqs.~(\ref{bdx1})~(\ref{xvbd}) with Eqs.~(\ref{Xp})~(\ref{Xvv}), one may find that the contributions of the ``images" to the correlation term in a flat spacetime with a boundary  are subtracted  from $X_0$ rather than added,  as is the case in the cosmic string spacetime.

Now let us  discuss what conclusions we can draw from Fig.~(\ref{CL1}). First, the presence of a cosmic string may either assist or inhibit entanglement harvesting,
depending on  the detector-to-string distance.  For a small detector-to-string distance ($l/\sigma\ll1$), the presence of the cosmic string will  assist entanglement harvesting in both parallel and orthogonal  alignments, which is in sharp contrast to the  inhibitory role played by the presence of a reflecting plane boundary in entanglement harvesting  in the vicinity of the boundary~\cite{Liu:2021,Liu:2023}. However, for a sufficiently large detector-to-string distance, given  a fixed interdetector separation $d$ and a fixed deficit-angle parameter $\nu$, the amount of  harvested entanglement  falls below the corresponding result in a flat spacetime without a cosmic string. This indicates that  the cosmic string acts as an inhibitor of entanglement harvesting, which contrasts with the assisting role played by a boundary in entanglement harvesting at sufficiently large detector-to-boundary distances.
Remarkably, the harvested entanglement with a cosmic string/boundary possesses a dip/peak at a certain large detector-to-string/detector-to-boundary distance.  Second,  Fig.~(\ref{CL1}) also shows that the detectors in the orthogonal-to-string/parallel-to-boundary alignment still have the potential to  harvest relatively more entanglement than the parallel-to-string/orthogonal-to-boundary alignment, provided that the detector-to-string/detector-to-boundary distance is sufficiently large, for a not-too-large interdetector separation $d$ and a fixed $\nu$.

\begin{figure}[!htbp]
\centering
\subfloat[$l/\sigma=0.10$ ]{\label{C-v(pv)11}\includegraphics[width=0.48\linewidth]{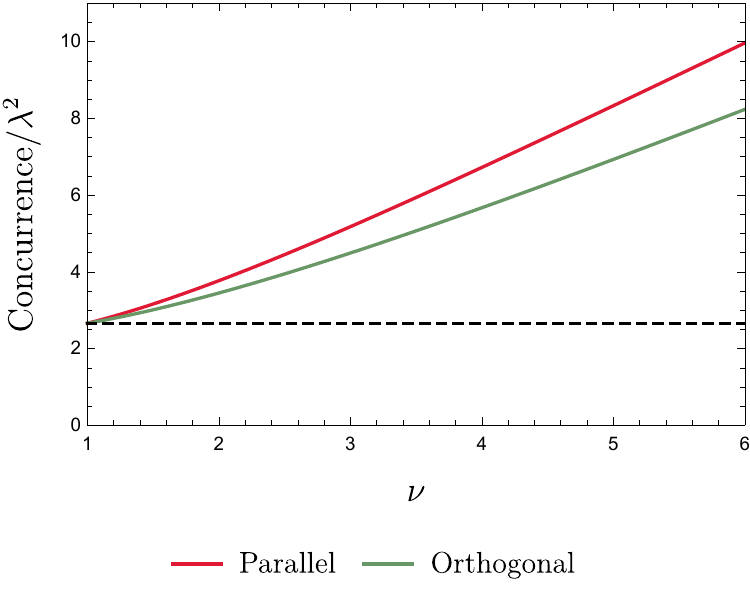}}~\;
\subfloat[$l/\sigma=3.00$]{\label{C-v(pv)13}\includegraphics[width=0.48\linewidth]{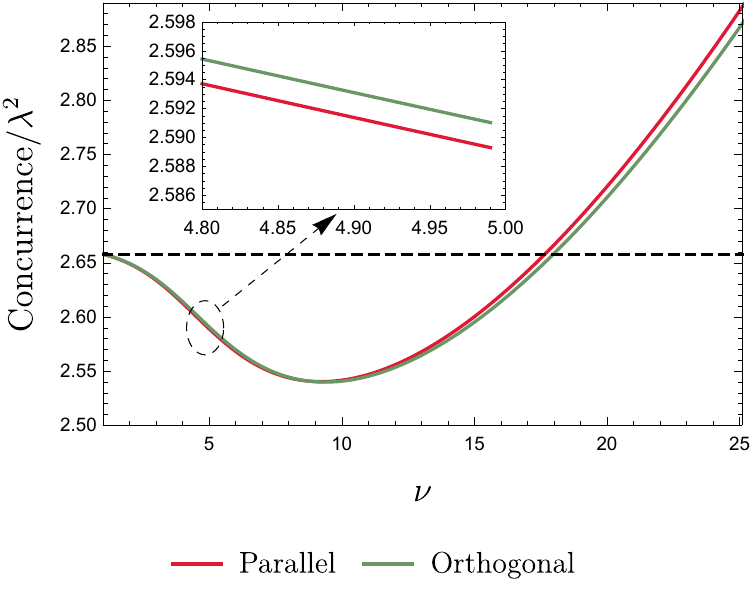}}\;
\caption{The concurrence is plotted as a function of $\nu$ for $l/\sigma=0.10$ in plot (a) and $l/\sigma=3.00$ in plot (b) with fixed $d/\sigma=0.10$ and $\Omega\sigma=0.10$. The black horizonal dashed line indicates the corresponding result in a flat spacetime without cosmic strings and boundaries.} \label{C-v(pv)}
\end{figure}

\begin{figure}[!htbp]\centering
\includegraphics[width=0.60\textwidth]{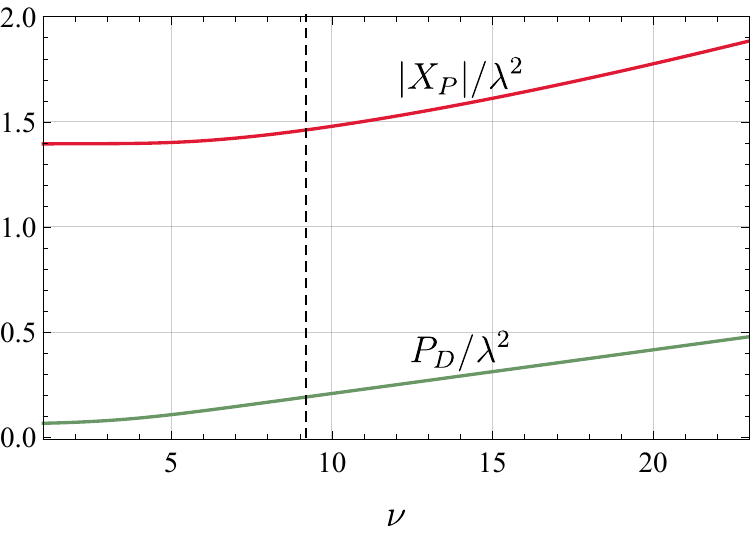}
\caption{ The correlation term $|X_{P}|$ and the transition probability $P_{D}$ are plotted as a function of parameter $\nu$ for $l/\sigma=3.00$, $\Omega\sigma=0.10$ and $d/\sigma=0.10$. The  dashed line ($\nu\approx9.220$) indicates where the minimum  difference between $|X_{P}|$ and $P_{D}$ occurs.}\label{xpvsv1}\end{figure}

In order to clearly reveal the influence of the deficit angle on  entanglement harvesting, we further plot the concurrence as a function of
the deficit-angle parameter in Fig.~(\ref{C-v(pv)}). Obviously, the  dependence of  the concurrence  on the deficit-angle parameter  is also significantly impacted by  the detector-to-string distance.  When the detector-to-string distance is small with respect to the duration time ($l/\sigma\ll1$), the concurrence is a monotonically increasing function of $\nu$, which is in accordance with our early analysis.  However, when the detector-to-string distance is not too small, the concurrence exhibits a dip at a special value of $\nu$, meaning it initially  decreases as $\nu$  increases, and subsequently becomes a continuously increasing function of $\nu$. To   understand this property clearly, we plotted the behaviors of $P_D$ and $|X_0|$ verses $\nu$  in the case of the parallel alignment in Fig.~(\ref{xpvsv1}).   As we can see,  both the transition probability $P_D$  and the  correlation term $|X_0|$ increase as $\nu$ grows, but their rates of increase differ.  Initially,  $P_D$  grows more rapidly than $|X_0|$ as  $\nu$ increases.  However, when $\nu$ reaches a sufficiently high value, the rate of increase in  $P_D$ no longer surpasses that of $|X_0|$. Given the formula for concurrence expressed as (\ref{Con}), we can infer that the concurrence would initially decrease and then begin to increase as $\nu$  continues to grow beyond a certain threshold. This behavior indicates a nuanced interplay between $P_D$ and $|X_0|$ that directly affects the  concurrence.

It is worth noting that  all the numerical results mentioned above are based on relatively short interdetector separations. Under these conditions, the detectors are causally connected, and as a result, the entanglement harvested does not solely originate from the vacuum. Instead, it is primarily generated through field-mediated communication between the two detectors. As argued in Ref.~\cite{Tjoa:2021}, entanglement harvesting includes two components: one stemming from the genuine entanglement preexisting in the vacuum, and the other from field-mediated communication between the detectors. When detectors are causally connected, the entanglement they acquire is predominantly influenced by this communication. In contrast, when the detectors remain spacelike separated, the harvested entanglement must necessarily arise solely from the preexisting vacuum entanglement.

\begin{figure}[!htbp]
\centering
\subfloat[$l/\sigma=0.10$]{\label{C-v(pv)31}\includegraphics[width=0.48\linewidth]{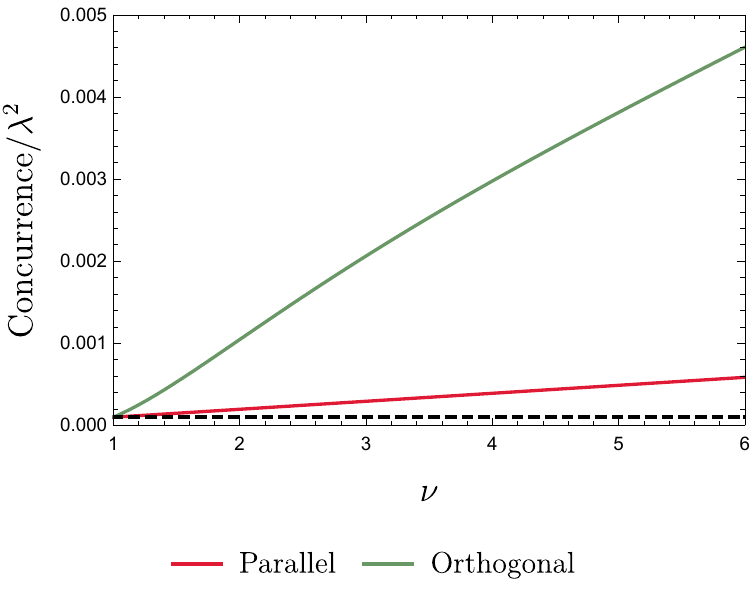}}\;
\subfloat[$l/\sigma=3.00$]{\label{C-v(pv)33}\includegraphics[width=0.48\linewidth]{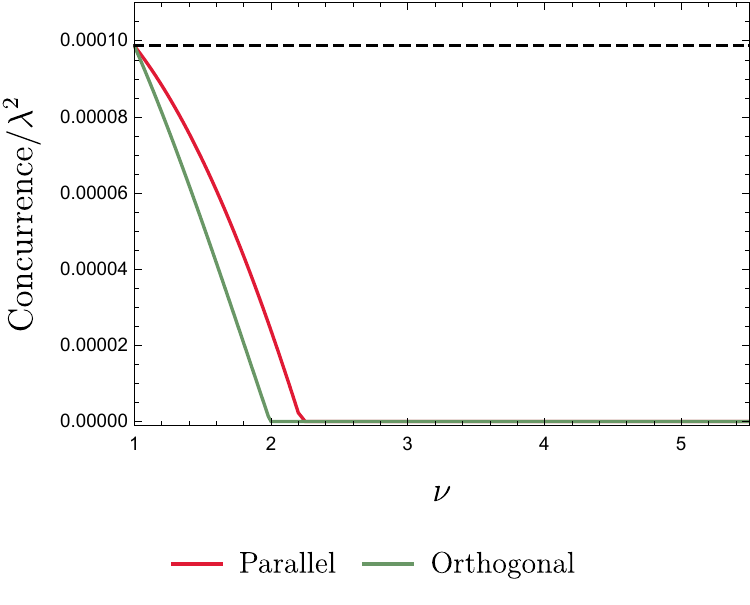}}
\caption{The concurrence is plotted as a function of $\nu$, for $l/\sigma=0.10$  in plot (a) and $l/\sigma=3.00$ in plot (b), with a relatively  large interdetector separation of $d/\sigma=4.00$ and $\Omega\sigma=1.50$ in the cases of parallel-to-string alignment and orthogonal-to-string alignment, respectively. Here, the black horizonal dashed line indicates the corresponding result in a flat spacetime without cosmic strings and boundaries.}\label{C-v-larged}
\end{figure}

To briefly explore how the presence of a cosmic string affects the genuine harvesting of preexisting entanglement in cosmic string spacetime, we have numerically estimated the concurrence acquired by two detectors placed at a large interdetector separation in Fig.~(\ref{C-v-larged}).  Here for a Gaussian switching function with a duration parameter $\sigma$, an interdetector separation $d=4\sigma$ is approximately regarded as effectively spacelike.  Thus, the contribution from communication between the detectors is expected to have a negligible impact on entanglement harvesting in this scenario, allowing the entanglement to be primarily harvested from the vacuum. This setup provides a clearer picture of the impact of cosmic strings on the genuine  entanglement harvesting from the vacuum.

As observed from Fig.~(\ref{C-v-larged}), the behavior of concurrence relative to the deficit-angle parameter $\nu$ varies with the detector-to-string distance. Specifically, when the distance between the detector and the string is small, the concurrence exhibits a monotonically increasing function of $\nu$. Conversely, if the detector-to-string distance is sufficiently large, the concurrence tends to show a monotonically decreasing trend as $\nu$ increases.

From these observations, one might infer that the influence of a cosmic string on genuine entanglement harvesting differs markedly depending on the proximity of the detectors to the string. In the vicinity of the string (the near zone), the presence of the cosmic string appears to enhance genuine entanglement harvesting, while at greater distances (the far zone), the string seems to inhibit this phenomenon compared to what is observed in a trivial flat spacetime. This distinction underscores the complex role that the topological defects like cosmic strings play in modifying the quantum field and the entanglement properties therein.

\begin{figure}[!htbp]\centering
\subfloat[$\Omega\sigma=0.10$]{\label{dmaxpv1}\includegraphics[width=0.48\linewidth]{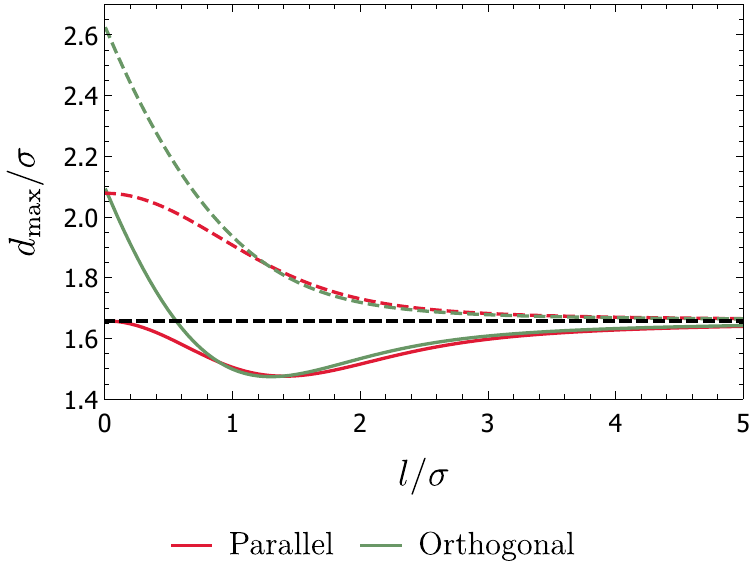}}\;
\subfloat[$\Omega\sigma=1.50$]{\label{dmaxpv2}\includegraphics[width=0.48\linewidth]{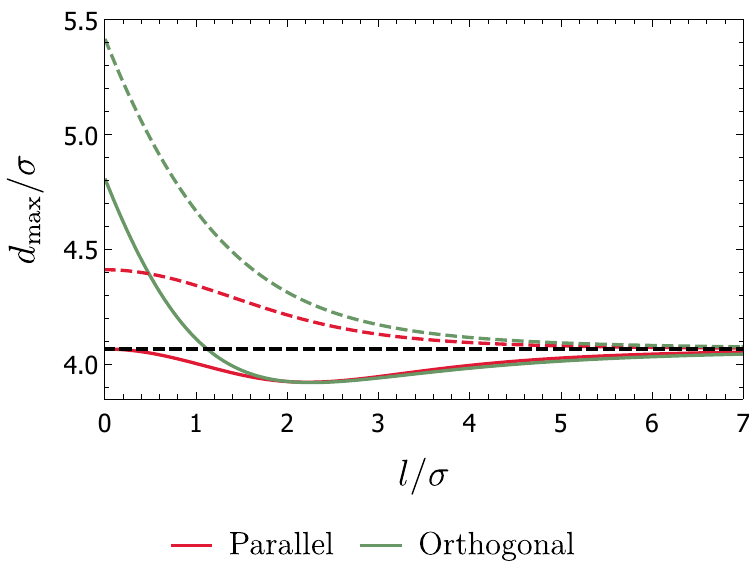}}
\caption{The maximum harvesting-achievable separation between detectors, $d_{\rm{max}}/\sigma$, is plotted as a function of $l/\sigma$ for the alignments with two detectors on the same side of the string with $\nu=3$ .\ Here, we have fixed $\Omega\sigma=0.10$ in (a) and $\Omega\sigma=1.50$ in (b). In all plots, the red dashed and green dashed lines respectively describe  the corresponding cases of parallel and orthogonal alignments in the presence of a reflecting boundary. The black dashed line  indicates the result in a flat spacetime without any strings or boundaries (a trivial flat spacetime).}\label{dmaxpv}
\end{figure}

Now we  analyze how the presence of a cosmic string impacts the harvesting-achievable range of interdetector separation.  We introduce  $d_{\rm{max}}$ to denote the maximum harvesting-achievable  separation, beyond which  entanglement harvesting cannot occur any more,  and plot it as a function of detector-to-string distance  in Fig.~(\ref{dmaxpv}).  As we can see from Fig.~(\ref{dmaxpv}), the presence of a cosmic string can either reduce or enlarge the harvesting-achievable range compared to the case of a trivial flat spacetime, depending on the detectors' alignment and the detector-to-string distance. Specifically, when the two detectors are aligned parallel to the string,  the presence of the cosmic string always reduces the harvesting-achievable range, sharply contrasting with the effect of a reflecting boundary, which always enlarges the range \cite{Liu:2021}.  However, when the detectors are aligned orthogonally to the string, the presence of the cosmic string tends to enlarge the harvesting-achievable range in the vicinity of the string. Yet, as the detector-to-string distance increases to become comparable to the duration parameter  ($l>\sigma$),
 it reduces the range. Thus, in terms of the harvesting-achievable range, the cosmic string plays a dual role in entanglement harvesting.

\subsection{ Orthogonal   alignment with two detectors  on two different sides of the string}

For this orthogonal  alignment [see Fig.~(\ref{model3})], the trajectories of  the detectors can be written as
\begin{equation}\label{sym-x}
x_{A}:= \left\{t=\tau,~\rho=l,~\theta=0,~z=0\right\},~x_{B}:= \left\{t=\tau,~\rho=d-l,~\theta=\pi,~z=0\right\}\;,
\end{equation}
where $l$ still represents the distance to the string of the detector which is closer and $d$ denotes the interdetector separation. Notice that now $d\geqslant2l>0$.
The transition probability $P_A$ is again Eq.~(\ref{pd}), and $P_B$ can now be obtained by replacing $l$ with $d-l$ in Eq.~(\ref{pd}).
Similarly, the nonlocal correlation term $X$, denoted here  by $X_T$ for the orthogonal  intersecting alignment, reads
\begin{equation}\label{Xs}
X_T=X_{0}+X_{T1}+X_{T2}\;,
\end{equation}
with
\begin{equation}\label{XT1}
X_{T1}=2\sum_{m=1}^{[\nu/2]}{'}f\Big(\sqrt{\frac{d^2}{4\sigma^2}-\frac{l(d-l)}{\sigma^2}\sin^{2}\frac{m\pi}{\nu}} \Big)\;,
\end{equation}
\begin{equation}\label{XT2}
X_{T2}=\int_{0}^{\infty}\mathrm{d}\zeta\frac{\nu\sin(2\nu\pi)}{2\pi[\cos(2\nu\pi)-\cosh(\nu\zeta)]}
f\Big(\sqrt{\frac{d^2}{4\sigma^2}+\frac{l(d-l)\cosh\zeta}{2\sigma^2}-\frac{l(d-l)}{2\sigma^2}} \Big)\;.
\end{equation}
It is easy to find out that both $X_{T1}$ and $X_{T2}$  vanish in the limit of $l\rightarrow\infty$, resulting in $X_T=X_0$.
For small $l/\sigma$ (i.e., $1\gg{l}/\sigma$ and $d\gg2l$), the correlation term  can then be approximated as
\begin{equation}\label{XTappr}
X_T\approx\left\{\begin{aligned}
&\nu{X_0},&{\text{integer}}~\nu\;,\\
&\Big(1+2\cdot\big[\frac{\nu}{2}\big]-\frac{1}{\pi}\operatorname{arctan~cot}({\nu\pi})\Big)X_0,&{\text{non-integer}}~\nu\;.
 \end{aligned} \right.
\end{equation}
And, for  small $d/\sigma$ and integer $\nu$, one may further have,   by taking Eqs.~(\ref{Con})~(\ref{plsmall}) and~(\ref{XTappr}) into account,  $\mathcal{C}_{T}(\rho_{A B})\approx\nu \mathcal{C}_{0}(\rho_{AB})$.

In particular, when the two detectors are in symmetric alignment with respect to the cosmic string (i.e., $d=2l$), we have
\begin{equation}\label{xT11}
X_{T1}=2\sum_{m=1}^{[\nu/2]}{'}f\Big(\frac{l}{\sigma}\cos\frac{m\pi}{\nu}\Big)\;,
\end{equation}
\begin{equation}\label{XT22}
X_{T2}=\int_{0}^{\infty}\mathrm{d}\zeta\frac{\nu\sin(2\nu\pi)}{2\pi[\cos(2\nu\pi)-\cosh(\nu\zeta)]}
f\Big(\frac{l}{\sigma}\cosh\frac{\zeta}{2} \Big)\;.
\end{equation}
When $\nu$ is an even integer, {$X_{T2}$ vanishes, and the last term ($m=\nu/2$) in the summation in Eq.~(\ref{xT11}) becomes
  $$f(0)=\lim_{z\rightarrow0}f(z)=\lim_{d\rightarrow0}X_0\rightarrow\infty\;,$$
 which consequently leads to a divergence in $X_{T}$ and,  accordingly, in the concurrence $\mathcal{C}_{T}(\rho_{A B})$ as well.
Physically,  this divergence results from the presence of an ``image" of detector $B$ that is angularly identical to detector $A$  due to the conical topology of the cosmic string spacetime, causing $A$  and $B$'s ``image" to spatially overlap. As mentioned previously, adopting a finite-size detector model with a spatial smearing function could resolve this divergence issue, although detailing such a model falls outside the main focus of this paper.  To clearly demonstrate  the properties of  entanglement harvesting, we will plot the concurrence as a function of the detector-to-string distance and the deficit-angle parameter in Figs.~(\ref{C-L(sym)}) and~(\ref{C-v(s)}), respectively.

\begin{figure}[!htbp]
\centering
\subfloat[$d/l=2.00$]{\label{C-L(d=2L)}\includegraphics[width=0.48\linewidth]{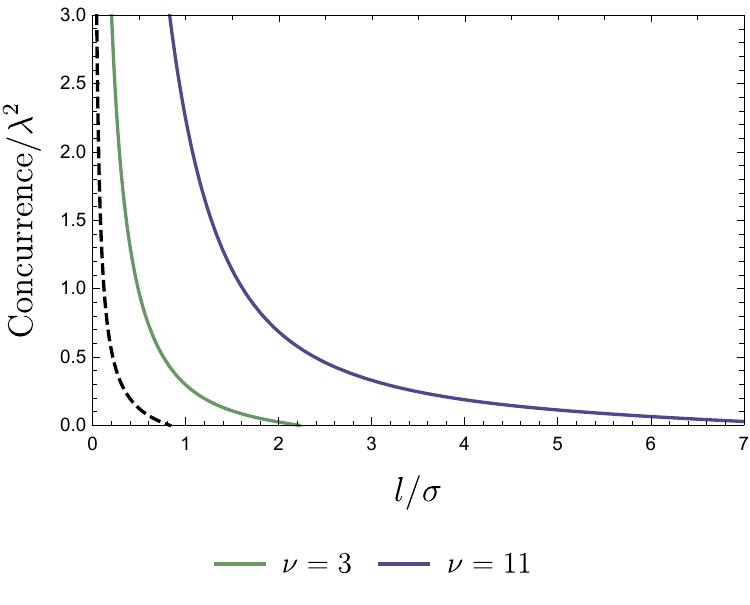}}\;
\subfloat[$d/l=2.50$]{\label{C-L(d=5-2L)}\includegraphics[width=0.48\linewidth]{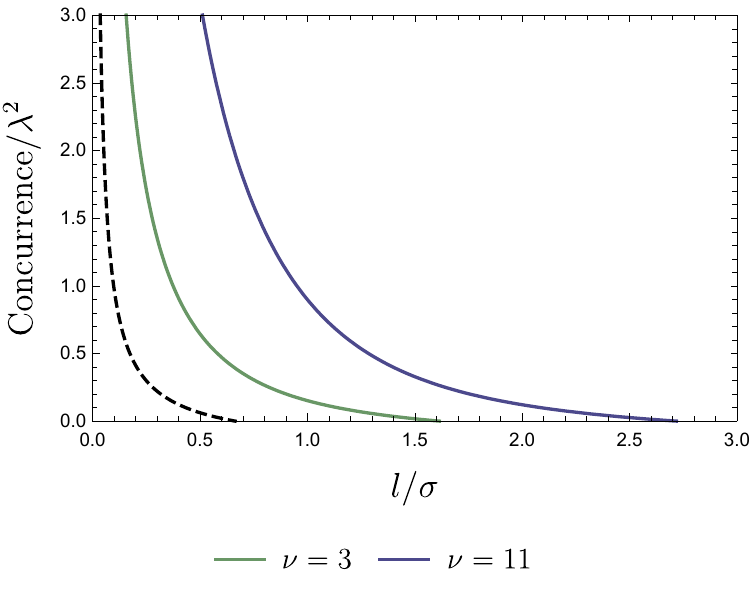}}
\caption{The concurrence is plotted  as a function of $l/\sigma$ with  $\Omega\sigma= 0.10$ for different interdetector separations $d/l=\{2.00,2.50\}$ in left-to-right order. The dashed lines represent the results in a flat spacetime without a cosmic string (i.e., $\nu=1$).}\label{C-L(sym)}
\end{figure}

As shown in Fig.~(\ref{C-L(sym)}), the concurrence is a monotonically decreasing function of the detector-to-string distance.
 This trend occurs because, in the orthogonal alignment with detectors on opposite sides of the string, the interdetector separation inherently increases with the detector-to-string distance, and concurrence typically decreases as the interdetector separation increases. Distinctly different from entanglement harvesting scenarios where detectors are positioned on the same side of the string, detectors aligned on opposite sides consistently harvest more entanglement compared to those in a flat spacetime without a cosmic string.

 \begin{figure}[!htbp]
\centering
\subfloat[$d/l=2.00, l/\sigma=0.10$]{\label{C-v(s)1}\includegraphics[width=0.46\linewidth]{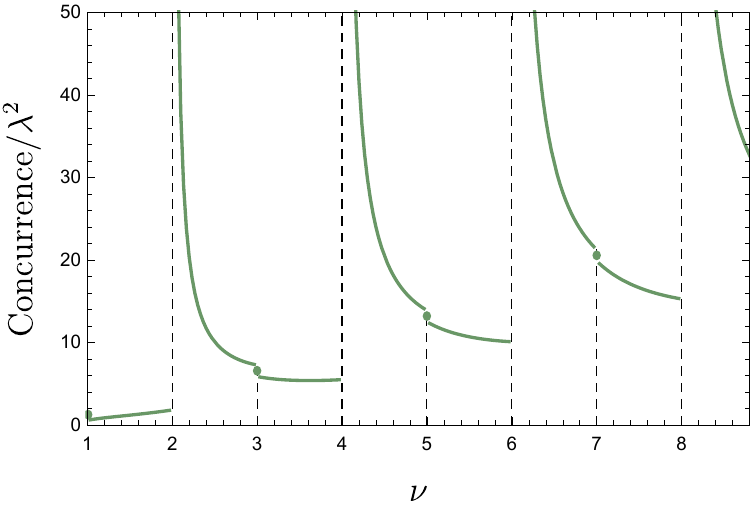}}\;
\subfloat[$d/l=2.00,l/\sigma=2.00$]{\label{C-v(s)2}\includegraphics[width=0.46\linewidth]{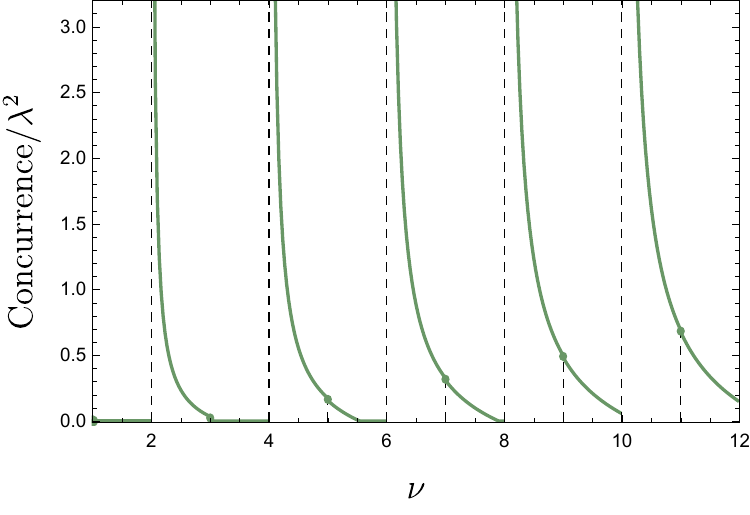}}\\
\subfloat[$d/l=2.50,l/\sigma=0.10$]{\label{C-v(ns)1}\includegraphics[width=0.46\linewidth]{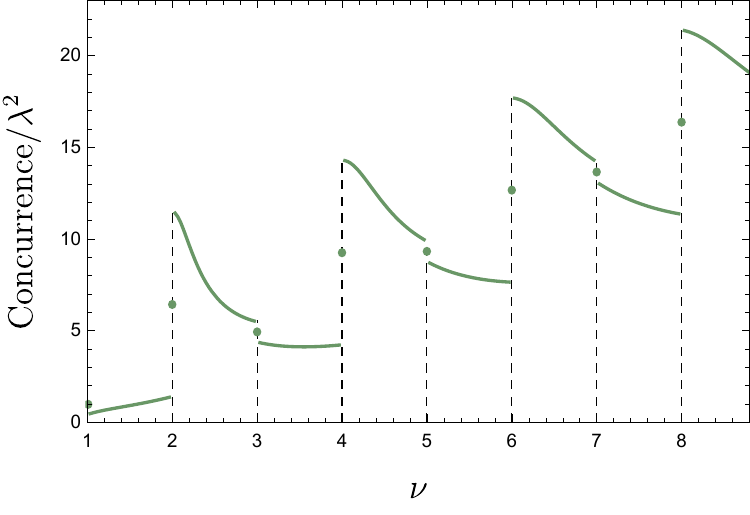}}\;
\subfloat[$d/l=2.50,l/\sigma=2.00$]{\label{C-v(ns)2}\includegraphics[width=0.46\linewidth]{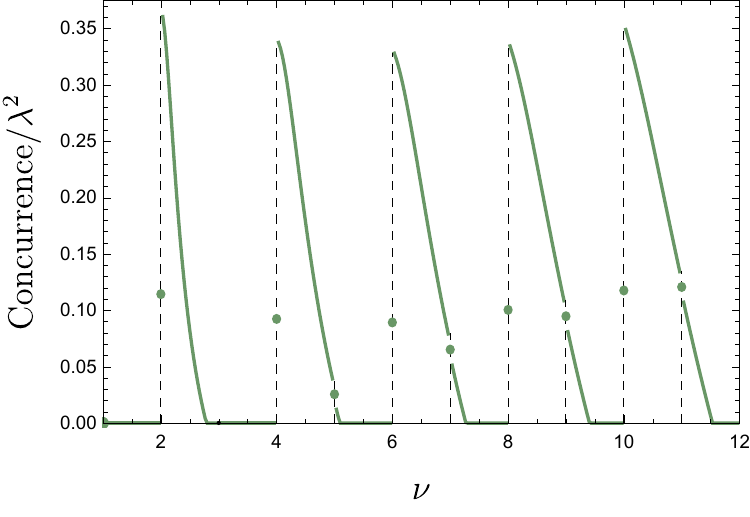}}
\caption{The concurrence  is plotted as a function of $\nu$ for  detectors in symmetrical alignment  with $l/\sigma=0.10$ in plot (a) and $l/\sigma=2.00$ in plot (b), and  for detectors  in non-symmetrical alignment with $d/l=2.50,\; l/\sigma=0.10$ in plot (c) and $d/l=2.50,\;l/\sigma=2.00$ in plot (d).  Here, we set $\Omega\sigma=0.10$ for all plots. The circle point exactly indicates the corresponding  value of concurrence at a certain integer $\nu$. }\label{C-v(s)}
\end{figure}

In Fig.(\ref{C-v(s)}), we explore how the concurrence varies with the deficit-angle parameter $\nu$. Unlike scenarios with both detectors on the same side of the string, the concurrence is no longer a continuous function at integer values of $\nu$, and  its dependence on $\nu$ does not vary qualitatively with the specific value of  $l/\sigma$.
  In particular, if the two detectors are  symmetrically aligned relative to the string ($d=2l$), the concurrence becomes divergent at even integer values of $\nu$. This divergence arises from the correlation term $X_T$, which becomes infinite due to the spatial overlapping of the detector and its image. This overlap leads to the breakdown of the point-like detector model [see plots~(\ref{C-v(s)1}) and~(\ref{C-v(s)2})]. Interestingly, there is a noticeable degradation in the harvested entanglement between every neighboring pair of even integers  $\nu$ [e.g., see  plots~(\ref{C-v(s)1}) and~(\ref{C-v(ns)1})]. Moreover, if $\nu$ is  an odd integer or even integer in cases of asymmetrical alignment with small interdetector separations,   the greater the value of $\nu$, the more the entanglement is harvested, indicating that the cosmic string indeed enhances entanglement harvesting.

  In addition, we plot  $d_{\rm{max}}/\sigma$   as a function of $l/\sigma$ in Fig.~(\ref{dmaxs}),  demonstrating that for not excessively large detector-to-string distances \footnote {Note that there exists a certain large  detector-to-string distance, beyond which  the detectors   cannot harvest entanglement any more due to the positive relationship between the interdetector separation and the detector-to-string distance ($d\geq2l$).},
   the harvesting-achievable range of interdetector separation with detectors on different sides of the string in cosmic string spacetime is consistently larger than that in trivial flat spacetime.   This observation suggests that, for this alignment, the impact of the cosmic string on the harvesting-achievable range is analogous to that of a reflecting boundary, facilitating enhanced entanglement harvesting under specific configurations.

\begin{figure}[!htbp]\centering
\includegraphics[width=0.60\textwidth]{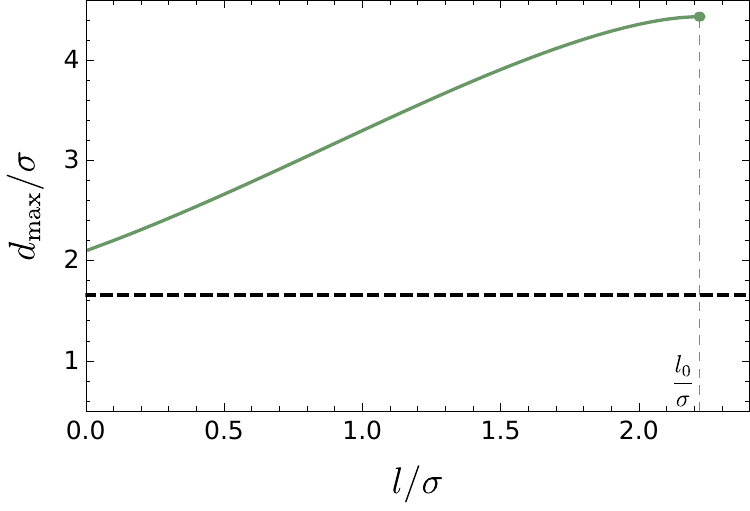}
\caption{The maximum harvesting-achievable interdetector separation versus the detector-to-string distance for the alignment with two detectors  on two different sides of the string. Here, we set $\nu=3$ and $\Omega\sigma=0.10$. The black dashed line represents the corresponding result in a flat spacetime without a cosmic string.  There is a certain value of the detector-to-string distance, i.e., $l=\l_0\approx2.219\sigma$, beyond which  the detectors   cannot harvest entanglement any more due to the positive relation between the interdetector separation and the detector-to-string distance ($d\geq2l$).}\label{dmaxs}
\end{figure}

\section{Conclusion}
\label{sec4}
In the framework of the entanglement harvesting protocol, we have performed a detailed study on the phenomenon of entanglement harvesting involving two Unruh-DeWitt (UDW) detectors that interact locally with a massless scalar field in the vicinity of a cosmic string. Specifically, we examined three different alignments of the detectors relative to the string: parallel, orthogonal with both detectors on the same side, and orthogonal with each detector on opposite sides. We find that the presence of the cosmic string in general  enhances the transition probability of the static detectors for a finite interaction duration.

For the alignments on the same side of the string,  we find the amount of entanglement harvested is always a monotonically decreasing function of  the interdetector separation, regardless of whether the  detectors are aligned parallel or orthogonally to the string. Meanwhile, the harvested entanglement does not  always decrease as the detector-to-string distance increases; instead, it displays a dip that  falls below the corresponding result in a flat spacetime without a cosmic string at a certain large detector-to-string distance. This behavior contrasts markedly with the case involving the presence of a reflecting boundary, where the harvested entanglement typically shows a peak as the detector-to-boundary distance grows to be comparable to the duration time parameter. In other words, the presence of a cosmic string may facilitate entanglement harvesting in its vicinity but tends to inhibit it in the far zone, which is the opposite of the effect observed with a boundary, where entanglement harvesting is inhibited near the boundary but assisted in the far zone.

Interestingly, when the detector-to-string distance is small relative to the interaction duration  parameter,  the larger the deficit angle,  the more the entanglement is harvested. However, if the detector-to-string distance is not too small with a not-too-large interdetector separation,  the entanglement harvested may initially  decrease slightly as the defect angle increases  but later becomes a monotonically increasing function as the deficit angle further enlarges.
Therefore, one may conclude that the amount of  entanglement harvested is generally  amplified due to the deficit angle, regardless of whether the two detectors are aligned parallel or orthogonally to the string, particularly when the deficit angle is sufficiently  large.    Notably, in the vicinity of the string,  the detectors in  parallel alignment with  a small interdetector separation can harvest more entanglement than those in orthogonal  alignment. However, the detectors in orthogonal alignment can, in turn, harvest comparatively more entanglement when the interdetector separation or the detector-to-string distance becomes sufficiently large. Moreover, the numerical results reveal that the presence of the cosmic string generally assists or inhibits genuine entanglement harvesting in the near or far zone of the cosmic string spacetime, respectively, in comparison with results in a flat spacetime.

As far as the harvesting-achievable range of interdetector separation is concerned, when the two detectors are aligned parallel to the cosmic string, the presence of the string always reduces the  harvesting-achievable range compared to the result in a trivial flat spacetime. However, when the detectors are aligned orthogonally to the string,  the harvesting-achievable range is enlarged in the vicinity of the string but  reduced in the far zone. Therefore, the presence of the cosmic string can either assist or inhibit entanglement harvesting in terms of the harvesting-achievable range, which markedly differs from the consistent enlargement of the harvesting-achievable range seen with a reflecting boundary.

Regarding the orthogonal  alignment with two detectors on  opposite  sides of the string, the amount of entanglement harvested always decreases as the detector-to-string distance increases, due to  the positive relationship between the detector-to-string distance and the interdetector separation. Remarkably,  unlike the scenario where both detectors are on the same side of the string, the detectors on opposite sides always harvest more entanglement than those in a trivial flat spacetime, and the presence of the cosmic string enlarges the harvesting-achievable range in the vicinity of the string.  Another interesting feature is that the harvested entanglement  is now a discontinuous  function of the deficit-angle parameter $\nu$, marking a significant deviation from the case with the two detectors on the same side.
 In particular, when the two detectors are  symmetrically aligned  with respect to the string ($d=2l$),  the harvested entanglement  diverges at  even integer values of $\nu$ due to the spatial overlapping of the detector and its image, a consequence of the conical topology of the cosmic string spacetime.

It is  worth noting that while the nontrivial topologies introduced by both the presence of a cosmic string and a reflecting boundary in a locally flat spacetime significantly impact the entanglement harvesting phenomenon, they exhibit distinctive influences. For instance, the cosmic string typically assists, in terms of the amount of entanglement harvested, the entanglement harvesting for two detectors on the same side in the vicinity and inhibits it in the far zone. Conversely, a boundary inhibits the entanglement harvesting in the vicinity and assists it in the far zone. However, in terms of the harvesting-achievable range, the cosmic string in the case of orthogonal alignment of the detectors relative to it plays a dual role, contrasting sharply with the consistent role of a reflecting boundary, which always enlarges the harvesting-achievable range. These sharply contrasting properties may provide a potential method to distinguish between locally flat spacetimes with different topologies due to cosmic strings and reflecting boundaries.

Finally, we have only investigated the entanglement harvesting for two detectors in three specific alignment cases, i.e., parallel, orthogonal on the same side, and orthogonal on opposite sides of the cosmic string.  However, one might be curious about scenarios where the detectors are not strictly parallel or orthogonal to the cosmic string. In fact, based on our analysis, when the detector system is positioned very far from the string, the entanglement harvesting phenomenon in the cosmic string spacetime behaves similarly to that in a trivial flat spacetime, which tends to be angle-independent. Thus, only in the vicinity of the string does the general configuration significantly impact entanglement harvesting. In scenarios where the angular difference between two detectors is close to 0 or $2\pi$ (i.e., $\cos(\theta_A-\theta_B)\sim1$), we can qualitatively infer that detectors with a small or large angle between the interdetector separation and the string axis may mimic the behavior of detectors in parallel or orthogonal alignment on the same side of the string, respectively. The presence of the string should enhance entanglement harvesting in terms of the amount of entanglement harvested in these misaligned cases, influencing both communication-mediated and genuine entanglement harvesting. Conversely, when the angular difference deviates significantly from 0 and  $2\pi$  the harvesting behavior diverges. Detectors with a large included angle between the interdetector separation and the string tend to harvest entanglement similarly to those in orthogonal alignment with detectors on opposite sides of the string. In contrast, a small included angle aligns more closely with parallel alignment harvesting characteristics. In these configurations, the influence of the cosmic string on entanglement harvesting in terms of the amount of entanglement harvested remains positive.
%%%%%%%%%%%%%%%%%%%%%%
\begin{acknowledgments}
This work was supported in part by the NSFC under Grants No.12175062 and No.12075084, and the Hunan Provincial Natural Science Foundation of China under Grant No. 2024JJ1006.
\end{acknowledgments}
%%%%%%%%%%%%%%%%%

\appendix
%%%%%%%%%%%%%%%%%%%%%%%%%%%%%%%
\section{The derivation of  $P_{D}$ }\label{appd1}
Let $u=\tau$ and $s=\tau-\tau'$,  and then the transition probability~(\ref{ppd}) can be rewritten as
\begin{align}\label{PAA1}
P_{D} & =\lambda^{2} \int_{-\infty}^{\infty} d u \chi_{D}(u) \int_{-\infty}^{\infty} d s \chi_{D}(u-s) e^{-i \Omega s} W(s) \nonumber\\
& =\lambda^{2} \sqrt{\pi} \sigma \int_{-\infty}^{\infty} d s e^{-i \Omega s} e^{-s^{2} /\left(4 \sigma^{2}\right)} W(s)\;.
\end{align}
Substituting the Wightman function~(\ref{WW2}) into Eq.~(\ref{PAA1}), the transition probability can be recast into a sum of three terms
\begin{equation}
P_{D}=P_{0}+P_{1}+P_{2}
\end{equation}
with
\begin{equation}\label{PA0}
P_{0}: =-\frac{\lambda^{2} \sigma}{4 \pi^{3 / 2}} \int_{-\infty}^{\infty} d{s} \frac{e^{-i\Omega s}e^{-\frac{s^{2}}{4\sigma^{2}}}}{(s-i\epsilon)^{2}}\;,
\end{equation}
\begin{equation}\label{PA1}
P_{1}:=-\frac{\lambda^{2} \sigma}{2 \pi^{3 / 2}}\sum_{m=1}^{[\nu/2]}{'} \int_{-\infty}^{\infty} d{s} \frac{e^{-i\Omega s}e^{-\frac{s^{2}}{4\sigma^{2}}}}{(s-i\epsilon)^{2}-4l^2\sin^2(\frac{m\pi}{\nu})}\;,
\end{equation}
\begin{equation}\label{PA2}
P_{2}: =\frac{\lambda^{2} \sigma\nu}{4 \pi^{5 / 2}}\int_{0}^{\infty}d{\zeta}\frac{\sin(\nu\pi)}{[\cosh(\nu\zeta)-\cos(\nu\pi)]} \int_{-\infty}^{\infty} d{s}\frac{ e^{-i\Omega s}e^{-\frac{s^{2}}{4\sigma^{2}}}}{(s-i\epsilon)^{2}-4l^2\cosh^2(\frac{\zeta}{2})}\;.
\end{equation}

All the integrals  above  can be calculated by using the technique of distribution functions.   Mathematically, the action of a distribution $g$ on a test
function $f$ is defined by
\begin{equation}
\langle{g},{f}\rangle:=\int_{-\infty}^{\infty}g(x)f(x)dx\;.
\end{equation}
There are some important identities for a distribution~\cite{Smith:2015,Bogolubov:1990}
 \begin{equation}\label{dist-r}
\big\langle{\frac{dg}{dx}},{f}\big\rangle=-\big\langle{g},{\frac{df}{dx}}\big\rangle\;.
\end{equation}
and
\begin{equation}\label{id1}
 \big\langle{\delta^{(n)}(x)},{f(x)}\big\rangle=(-1)^nf^{(n)}(0)\;.
 \end{equation}
 Especially for the distributions $1/x$ and $1/x^2$, we have
\begin{equation}\label{id2}
\big\langle{\frac{1}{x}},{f(x)}\big\rangle={\cal{P}}\int_{-\infty}^{\infty}\frac{f(x)}{x}dx\;,
\end{equation}
\begin{equation}\label{id3}
\big\langle{\frac{1}{x^2}},{f(x)}\big\rangle=\int_{0}^{\infty}dx\frac{f(x)+f(-x)-2f(0)}{x^2}\;,
\end{equation}
where $\cal{P}$ denotes the principle value of an integral.  The first integral for $P_0$ can be evaluated  by using  the following identity that arises from differentiation of the Sokhotski-Plemelj formula
\begin{equation}\label{id4}
\frac{1}{(x \pm i \epsilon)^{n}}=\frac{1}{x^{n}} \pm \frac{(-1)^{n}}{(n-1) !} i \pi \delta^{(n-1)}(x)\;.
\end{equation}
So, with these identities it is easy to obtain that~\cite{Smith:2015}
\begin{align}
P_{0} & =-\frac{\lambda^{2} \sigma}{4 \pi^{3 / 2}} \int_{-\infty}^{\infty} d{s} e^{-i\Omega s}e^{-s^{2}/(4\sigma^{2})}\frac{1}{(s-i\epsilon)^{2}}\nonumber\\
& =-\frac{\lambda^{2} \sigma}{4 \pi^{3 / 2}} \int_{-\infty}^{\infty} d{s} \frac{e^{-i \Omega s}e^{-s^{2} / (4 \sigma^{2})}}{s^{2}}+\frac{i\lambda^{2} \sigma}{4 \sqrt{\pi}} \int_{-\infty}^{\infty} d{s} e^{-i \Omega s}e^{-s^{2} / (4 \sigma^{2})}\delta^{(1)}(s)\nonumber\\& =-\frac{\lambda^{2} \sigma}{4 \pi^{3 / 2}} \int_{0}^{\infty} d{s} \frac{e^{-i \Omega s}e^{-s^{2} / (4 \sigma^{2})}+e^{i \Omega s}e^{-s^{2} / (4 \sigma^{2})}-2}{s^{2}}-\frac{\lambda^{2} \Omega\sigma}{4 \sqrt{\pi}}\nonumber\\
& =\frac{\lambda^{2}}{4\pi}\left[e^{-\sigma^{2}\Omega^{2}}-\sqrt{\pi}\sigma\Omega \operatorname{Erfc}(\sigma \Omega)\right].
\end{align}
With the help of the identity~(\ref{id4}), the second term of the transition probability  can be written as
\begin{align}
P_{1}=& -\frac{i\lambda^{2} \sigma}{2 \sqrt{\pi}}\sum_{m = 1}^{[\nu/2]}{'} \int_{-\infty}^{\infty} d{s} e^{-i\Omega s}e^{-\frac{s^{2}}{4\sigma^{2}}}{\rm{sgn}}(s)\delta\Big(s^{2}-4l^{2}\sin^{2}\frac{m\pi}{\nu}\Big)\nonumber\\
&-\frac{\lambda^{2} \sigma}{2 \pi^{3 / 2}} \sum_{m = 1}^{[\nu/2]}{'}\;{\cal{P}}\int_{-\infty}^{\infty} d{s} e^{-i\Omega s}e^{-\frac{s^{2}}{4\sigma^{2}}}\frac{1}{s^{2}-4l^2\sin^2\frac{m\pi }{\nu}}\nonumber\\
=&-\frac{\lambda^{2} \sigma}{4l \sqrt{\pi}}\sum_{m = 1}^{[\nu/2]}{'}\;\frac{e^{-l^{2}\sin^{2}({m\pi }/{\nu})/ \sigma^{2}}}{\sin({m\pi}/{\nu})}  \sin \Big[2 l\Omega \sin\big({m\pi }/{\nu}\big)\Big]\nonumber\\&-\frac{\lambda^{2} \sigma}{2 \pi^{3 / 2}} \sum_{m = 1}^{[\nu/2]}{'}\;{\cal{P}}\int_{-\infty}^{\infty} d{s} e^{-i\Omega s}\frac{e^{-s^{2}/(4\sigma^{2})}}{s^{2}-4l^2\sin^2\frac{m\pi }{\nu}}\nonumber\\
=&\frac{\lambda^{2} \sigma}{4l \sqrt{\pi}}\sum_{m=1}^{[\nu/2]}{'}\;\frac{e^{-l^{2}\sin^{2}({m\pi}/{\nu})/\sigma^{2}}}{\sin({m\pi}/{\nu})}\bigg\{\operatorname{Im}\Big[e^{2 i \Omega l\sin({m\pi}/{\nu})} \operatorname{Erf}\Big(\frac{il}{\sigma}\sin\frac{m\pi}{\nu}+\sigma\Omega\Big)\Big]\nonumber\\
&-\sin \Big[2l\Omega \sin\big({m\pi }/{\nu}\big)\Big]\bigg\}\;,
\end{align}
where we have performed the integration by using the convolution of two functions in the Fourier transforms in the last step.

Similarly, the  integration with respect to $s$ in Eq.~(\ref{PA2}) can also be straightforwardly  carried out firstly, and after some  manipulations, one has
\begin{align}
P_{2}=&\frac{\lambda^{2} \sigma\nu\sin(\nu\pi)}{8l \pi^{3/2}}\int_{0}^{\infty} d\zeta\frac{1}{\cosh(\nu\zeta)-\cos(\nu\pi)} \frac{e^{-l^{2} \cosh^{2}({\zeta}/{2})/\sigma^{2}}}{ \cosh({\zeta}/{2})}\bigg\{\sin\big[2l\Omega \cosh({\zeta}/{2})\big]\nonumber\\
&-\operatorname{Im}\Big[e^{2 i l\Omega  \cosh({\zeta}/{2})} \operatorname{Erf}\Big(\frac{{i}l}{\sigma}\cosh\frac{\zeta}{2}+\sigma \Omega\Big)\Big]\bigg\}\;.
\end{align}

\section{The derivation of  $X_P$ }\label{appd2}

Substituting the trajectories~(\ref{para-x}) into Eq.~(\ref{wxx}), one finds
\begin{align}\label{B1}
W&\big(x_{A}(\tau), x_{B}(\tau')\big)=W\big(x_{B}(\tau), x_{A}(\tau')\big)\nonumber\\
=& -\frac{1}{4 \pi^{2}} \frac{1}{(\tau-\tau'-i\epsilon)^2-d^{2}}-\frac{1}{2 \pi^{2}} \sum_{m=1}^{[\nu/2]}{'} \frac{1}{(\tau-\tau'-i\epsilon)^2-d^{2}-4l^2\sin^2(\frac{m\pi}{\nu})} \nonumber\\
& +\frac{\nu}{4 \pi^{3}} \int_{0}^{\infty} d \zeta \frac{\sin (\nu\pi)}{\big[\cosh (\nu \zeta)-\cos (\nu \pi)\big]\big[(\tau-\tau'-i\epsilon)^2-d^{2}-2l^{2}(1+\cosh\zeta)\big]}\;.
\end{align}

From Eq.~(\ref{X}), after assuming $u:=\tau'-\tau$, $u':=\tau'+\tau$, we have
 \begin{align}\label{B2}
X_{P}& = -\lambda^{2} \int_{-\infty}^{\infty} d u^{\prime}e^{-u'^{2} / (4 \sigma^{2})}e^{-i \Omega u'} \int_{0}^{\infty} d u e^{-u^{2} / (4 \sigma^{2})}W\big(u)\nonumber\\
&=-2 \sqrt{\pi} \lambda^{2} \sigma e^{-\sigma^{2} \Omega^{2}}\int_{0}^{\infty} d u e^{-u^{2} / (4 \sigma^{2})}W\big(u)\nonumber\\
&=X_{0}+X_{P1}+X_{P2}\;
\end{align}
with
\begin{equation}\label{xB0}
X_{0}: = \frac{  \lambda^{2} \sigma}{2\sqrt{\pi^3}} e^{-\sigma^{2} \Omega^{2}}\int_{0}^{\infty} d u  \frac{e^{-u^{2}/(4 \sigma^{2})}}{(-u-i\epsilon)^2-d^{2}}\;,
\end{equation}
\begin{equation}\label{xB1}
X_{P1}: = \frac{ \lambda^{2} \sigma}{\sqrt{\pi^3}}e^{-\sigma^{2} \Omega^{2}}\sum_{m=1}^{[\nu/2]}{'} \int_{0}^{\infty} d u  \frac{e^{-u^{2}/(4 \sigma^{2})}}{(-u-i\epsilon)^2-d^{2}-4l^2\sin^2({m\pi }/{\nu})}\;,
\end{equation}
\begin{equation}\label{xB2}
X_{P2}: = -\frac{ \nu\lambda^{2} \sigma}{2\sqrt{\pi^5}}{e}^{-\sigma^{2} \Omega^{2}}\int_{0}^{\infty} d\zeta \frac{\sin (\nu\pi)}{\cosh (\nu \zeta)-\cos (\nu\pi)}\int_{0}^{\infty} d u  \frac{e^{-u^{2}/(4 \sigma^{2})}}{(-u-i\epsilon)^2-d^{2}-2l^{2}(1+\cosh\zeta)}\;.
\end{equation}
According to the identity.~(\ref{id4}), $X_0$ can be worked out directly~\cite{Smith:2015}
\begin{align}
X_{0} & =\frac{  \lambda^{2} \sigma}{2\sqrt{\pi^3}} e^{-\sigma^{2} \Omega^{2}}{\cal{P}} \int_{0}^{\infty} du \frac{e^{-u^{2}/(4 \sigma^{2})}}{u^2-d^{2}}- \frac{i\lambda^{2}\sigma}{4d \sqrt{\pi}}  e^{-\sigma^{2} \Omega^{2}-d^{2} /(4\sigma^{2})} \nonumber\\
& =- \frac{i\lambda^{2}\sigma}{4 d\sqrt{\pi}} e^{-\sigma^{2} \Omega^{2}-d^{2} / (4 \sigma^{2})}\operatorname{Erfc}\Big( \frac{i d}{2 \sigma}\Big)=f\Big(\frac{d}{2\sigma}\Big)\;,
\end{align}
 with the auxiliary function $f(z)$ is defined by Eq.~(\ref{fzd}).
 Similarly, after carrying out the integration with respect to $u$ in Eqs.~(\ref{xB1} ) and~(\ref{xB2}),  we have
 \begin{equation}
X_{P1}=2\sum_{m=1}^{[\nu/2]}{'}f\Big(\sqrt{\frac{d^2}{4\sigma^2}+\frac{l^2}{\sigma^2}\sin^{2}\frac{m\pi}{\nu}} \Big)\;,
\end{equation}
and
\begin{equation}
X_{P2}=\int_{0}^{\infty}\mathrm{d}\zeta\frac{\nu\sin(\nu\pi)}{\pi[\cos(\nu\pi)-\cosh(\nu\zeta)]}
f\Big(\sqrt{\frac{d^2}{4\sigma^2}+\frac{l^2}{2\sigma^2}+\frac{l^2\cosh\zeta}{2\sigma^2}} \Big)\;.
\end{equation}

\end{document}